\pdfoutput=1
\documentclass[aps,prd,reprint,superscriptaddress,longbibliography,nofootinbib]{revtex4-2}
\usepackage{graphicx}
\usepackage{amsmath}
\usepackage{blindtext}
\usepackage{amssymb}
\usepackage{amsfonts}
\usepackage{dcolumn}
\usepackage{dsfont}
\usepackage{makecell}
\usepackage{latexsym}
\usepackage{rotating}
\usepackage{color}
\usepackage{amssymb}
\usepackage{pifont}
\usepackage{latexsym}
\usepackage{bbm}
\usepackage{subfigure}
\usepackage{float}
\usepackage{epsfig}
\usepackage{psfrag}
\usepackage{natbib, hyperref}
\usepackage{bm}
\usepackage{amsthm}
\usepackage[normalem]{ulem}
\usepackage{eucal}
\usepackage{mathrsfs}
\usepackage{url}
\usepackage{braket}
\usepackage{multirow}
\usepackage{ulem}
\usepackage{calligra}
\usepackage[utf8]{inputenc}
\usepackage{newunicodechar}
\usepackage{marvosym}
\usepackage[absolute]{textpos}
\usepackage[cal=cm]{mathalfa}
\usepackage{soul}
\usepackage{gensymb}

\usepackage{color}

\usepackage{hyperref}
\hypersetup{
colorlinks=true,final=true,
        linkcolor=blue,
        citecolor=blue,
        filecolor=blue,
        urlcolor=blue,
}

\begin{document}
\setlength{\abovedisplayskip}{3pt}
\setlength{\belowdisplayskip}{3pt}

\title{Interface controlled Berry phase and anisotropic spin–charge conversion in 
altermagnet–topological insulator bilayers}

\author{Juhi Singh}
\email{juhi_s@ph.iitr.ac.in}
\affiliation{Department of Physics, Indian Institute of Technology Roorkee, Roorkee 247667, India}
\author{Narayan Mohanta}
\email{narayan.mohanta@ph.iitr.ac.in}
\affiliation{Department of Physics, Indian Institute of Technology Roorkee, Roorkee 247667, India}

\begin{abstract}
We propose an altermagnet–topological insulator bilayer as a platform to engineer Berry phase driven spin–charge responses using an interfacial buffer layer. Using a momentum-space lattice model and linear-response theory, we investigate a $d$-wave altermagnet coupled to a topological insulator and highlight the crucial role of spin-flip tunneling in shaping its electronic and transport properties. Interfacial hybridization strongly modifies the band structure, leading to anisotropic Rashba–Edelstein and Hall responses. The spin-flip component of the coupling induces an inverse $d$-wave spin texture in the altermagnetic bands, signaling the onset of an altermagnetic topological phase. This coupling also renders the Rashba–Edelstein effect strongly in-plane anisotropic, enhancing the transverse response relative to ferromagnetic or antiferromagnetic analogues. These results establish interfacial spin-flip tunneling as a practical control knob for direction-sensitive, stray-field–free spin–charge conversion in correlated topological heterostructures.
\end{abstract}

\maketitle

\maketitle

\section{Introduction}

The recent discovery of altermagnets (AMs) has established a new class of magnetic order beyond ferromagnetism and antiferromagnetism, characterized by momentum-dependent spin splitting and a vanishing net magnetization \cite{Smejkal_PRX2022b, Smejkal_PRX2022a, Smejkal_NatRevMater2022}. This symmetry-protected order has been linked to unconventional transport responses, including anomalous Hall effects reported in both theoretical and experimental studies \cite{Attias_PRB2024,Helena_NatCommun2024,Feng_NatElectron2022,Sheoran_PRB2025,Yu_npjQuantumMater2025,Smejkal_SciAdv2020,Farajollahpour_npj2025,Bo_PRB2024,Hariki_PRB2023}. Beyond Hall transport, recent works have uncovered a variety of altermagnetic topological phases, including nontrivial Chern states, higher-order topological phases, and superconducting realizations \cite{Gonzalez_PRB2025,Yu-Xuan_PRB2024,Brekke_PRB2023,RuiChen_PRL2025}. Collectively, these advances establish AMs as a fertile ground for realizing symmetry-driven responses distinct from those found in conventional magnets. In parallel, topological insulators (TIs) have been at the forefront of condensed matter research in the past decade, due to their spin-momentum–locked surface states, which enable dissipationless transport and efficient spin–charge interconversion \cite{Hasan_RMP2010,Bernevig_Science2006, Qi_Zhang_RevModPhys2011, Hsieh_Nature2009, Shi_PRB2025, Nagaosa_AHE_RMP2010, Chang_RevModPhy2023}. Mechanisms such as the Rashba–Edelstein effect and inverse Edelstein effect provide key routes for charge-to-spin conversion \cite{Rashba_1984,Edelstein_SSC1990,Rojas_NatCommun2013,Lesne_NatMater2016}. Spin pumping and spin–torque experiments have further demonstrated that TIs and Rashba interfaces are highly effective spin sources \cite{Mellnik_Nature2014,Liu_NatCommu2018,Shiomi_PRL2014}. These phenomena establish TI as a versatile platform for engineering spin responses in hybrid systems.

When magnetic systems are interfaced with spin-orbit coupled materials, interlayer hybridization can produce novel transport effects by mixing spin and orbital degrees of freedom \cite{Felser_NatMater2022, MacNeill_PRL2017, Manchon_RMP2019, Seibold_PRL2017, Guillet_PRB2020, Kondou_NatPhys2016, DeyRik_PRB2018, Haoran_PRB2021, Leiva_PRevRes2023, Annika_PRB2016} . In ferromagnet–TI (FM-TI) and antiferromagnet–TI (AFM-TI) bilayers, interfacial couplings have been shown to enable spin injection, torque generation, and spin–charge conversion \cite{Fan_NatMater2014, Zelezny_PRL2014, Arati_PRB2016, Lv_NatCommun2018, Lv_ApplPhysRev2022, JSingh_PRB2024, Dhavala_PRB2024, Mohanta_SciRep2017, Wang_PRL2015, Garate_PRL2010, Chen_PRB2020, Ghosh_PRB2018}. However, these responses are constrained by net magnetization or symmetry restrictions. Altermagnets, by contrast, combine momentum-dependent spin splitting with vanishing magnetization, offering a distinct route to anisotropic spintronic effects without stray fields. Recent theoretical studies of AM–TI interfaces have identified Néel-vector–controlled topological transitions \cite{Ezawa_PRB2025}, selection rules governing altermagnetic topological insulators \cite{Hai-Yang_PRB2024}, emergent spin–valley locking driven by crystal symmetries \cite{Ma_Quantumfront2024}. These findings highlight the rich phenomenology possible in AM–TI systems and motivate a systematic study of their spin textures and transport responses.

\begin{figure}[b]
\begin{center}
\vspace{-0mm}
\epsfig{file=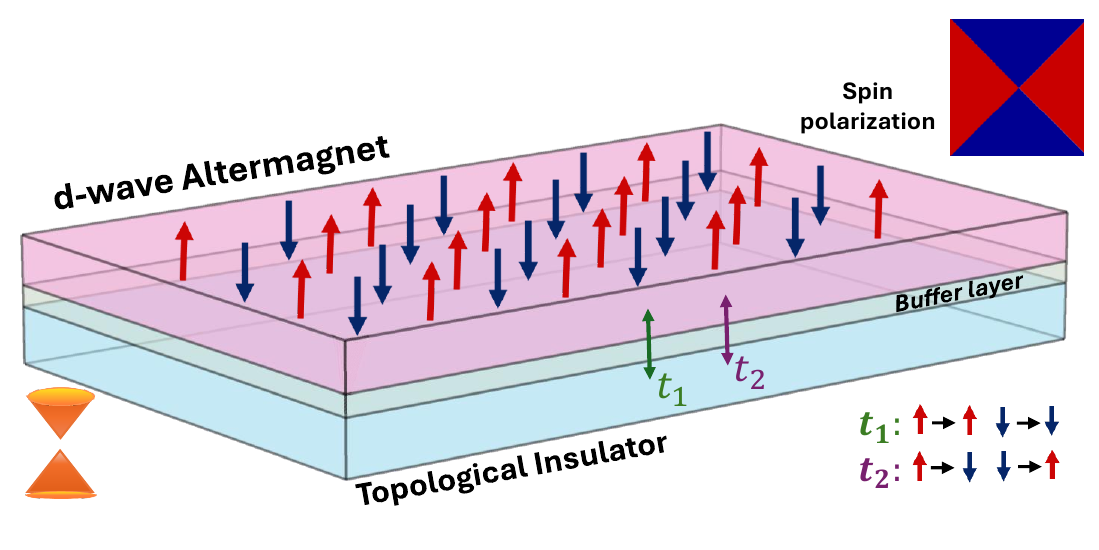,trim=0.0in 0.0in 0.0in 0.0in,clip=false, width=86mm}
\caption{Schematic of the two-dimensional bilayer: a $d$-wave altermagnet (top) stacked on a topological insulator surface (bottom). Interlayer hybridization proceeds via a spin-conserving hopping channel $t_{1}$ and a spin-flip channel $t_{2}$, which couple same-spin and opposite-spin states across the interface, respectively.}
\label{fig1}
\vspace{-4mm}
\end{center}
\end{figure} 

In this work, we investigate bilayer heterostructures comprising a $d$-wave altermagnet and a topological insulator, and show that interfacial hybridization reshapes both the electronic structure and transport properties. The system geometry is illustrated in Fig.~\ref{fig1}, a two-dimensional altermagnet layer is stacked atop the topological-insulator surface, with interlayer coupling proceeding through a spin-conserving hopping channel $t_{1}$ and a spin-flip channel $t_{2}$. The $t_{1}$ term connects states with the same spin orientation across the interface, whereas $t_{2}$ mixes opposite-spin components, effectively acting as an interfacial spin–orbit field. Using a momentum-space lattice model and linear-response theory, we demonstrate that this hybridization modifies the band topology, producing anisotropic Rashba–Edelstein and Hall responses. The spin-flip channel, in particular, induces an inverse $d$-wave spin texture in the altermagnetic bands, signaling the emergence of an altermagnetic topological phase. These hybridized spin configurations are qualitatively distinct from those of ferromagnetic or antiferromagnetic counterparts and manifest in transport signatures that are highly sensitive to crystalline symmetry. Our analysis further reveals enhanced anisotropy in Edelstein and Hall responses compared to conventional magnetic materials, establishing AM–TI bilayers as a new platform for amplifying spin–charge interconversion.

An important feature of this proposal is the tunability of the interfacial coupling. 
The strength of spin-conserving and spin-flip tunneling can be experimentally engineered by introducing buffer layers of varying thicknesses or materials with strong spin–orbit interactions. External electric fields can also tune these tunneling parameters by modifying the interfacial potential gradient, which changes the effective overlap between the altermagnet and topological insulator states \cite{Hui_SciRep2015}.
Such interfacial control has been demonstrated in TI-based heterostructures where giant spin–orbit torques drive magnetization switching \cite{Fan_NatMater2014}, in antiferromagnets where relativistic SOC produces current-induced Néel-order fields \cite{Zelezny_PRL2014}, and in epitaxial AFM thin films showing tunable spin Seebeck effects \cite{Arati_PRB2016}. Together, these works confirm that interfacial SOC and magnetic order can be practically manipulated, reinforcing the feasibility of controlling anisotropic spin–charge conversion in AM–TI heterostructures.

\section{Model and Method}

We model the AM-TI bilayer with a $d$-wave AM stacked on a TI, as shown in Fig.~\ref{fig1}. The effective low-energy description is constructed in momentum space and incorporates the altermagnetic order, the Rashba-coupled TI surface state, and interlayer tunneling that includes both spin-conserving and spin-flip channels. 
The Hamiltonian for this bilayer is given by

\begin{equation}
  H(\mathbf{k}) = \begin{bmatrix} \mathcal{H}_{\text{AM}}(\mathbf{k}) & H_{\text{int}} \\ H^{\dagger}_{\text{int}} & \mathcal{H}_{\text{TI}}(\mathbf{k}) \end{bmatrix}, \label{H_total}
\end{equation}

constructed in the basis $(d_{\mathbf{k},\uparrow}\, , \, d_{\mathbf{k}, \downarrow}\, , \, c_{\mathbf{k}, \uparrow}\, , \, c_{\mathbf{k}, \downarrow} )^{T}$, where $d_{\mathbf{k}, \sigma}$ and $c_{\mathbf{k}, \sigma}$ denote electron annihilation operators with spin $\sigma \in \{\uparrow, \downarrow \}$ in the AM and TI layers, respectively. In the above equation, $H_{\text{AM}}(\mathbf{k})$ represents the Hamiltonian for the $d$-wave AM with its N\'eel vector along the $z$ direction, expressed as

\begin{align}
\mathcal{H}_{\text{AM}} (\mathbf{k}) =  \sum_{\sigma,\sigma'} 
d^{\dagger}_{\mathbf{k},\sigma} \left[H_{\text{AM}}(\mathbf{k})\right]_{\sigma\sigma'} 
d_{\mathbf{k},\sigma'}, 
\label{H_AM}
\end{align}

where $H_{\text{AM}}(\mathbf{k})= 2 t_J (\cos k_x - \cos k_y) \sigma_z$, $t_J$ is the spin-dependent hopping amplitude due to anisotropic exchange coupling, $\boldsymbol{\sigma}=(\sigma_x,\sigma_y,\sigma_z)$ denotes Pauli matrices in spin space, and $\mathbf{k}\equiv (k_x, k_y)$ represents the two-dimensional momentum at the interfacial plane. This form captures the essential symmetry of $d$-wave altermagnetism—momentum-dependent spin splitting with vanishing net magnetization\cite{Smejkal_PRX2022b, Smejkal_PRX2022a}. It is interesting to note that two-dimensional AMs can exhibit spin-orbit driven splitting at the $\Gamma$ point and weak ferromagnetism \cite{Joachim_APL2024, Milivojević_2DMater2024}. Our model excludes intrinsic spin-orbit coupling in AMs to focus on the proximity effect originating from the topological insulator surface state. Since the N\'eel vector in the above Hamiltonian is along the $z$ direction, the AM preserves two-fold rotation ($C_{2z}$) and vertical mirror symmetries ($M_x$, $M_y$) but breaks time reversal ($\Theta$). It can, however, retain a combined four-fold rotation–time reversal symmetry ($C_{4z}\Theta$) that enforces degenerate spin pairs at certain high-symmetry points. 

The TI surface is modeled by the following Rashba-like Hamiltonian
\begin{align}
\mathcal{H}_{\mathrm{TI}} (\mathbf{k}) = \sum_{\sigma,\sigma'} 
c^{\dagger}_{\mathbf{k},\sigma} [H_{\mathrm{TI}}(\mathbf{k})]_{\sigma\sigma'} c_{\mathbf{k},\sigma'},
\label{H_TI}
\end{align}

where $H_{\mathrm{TI}}(\mathbf{k}) = \lambda_R \left(\sin k_y\,\sigma_x - \sin k_x\,\sigma_y\right)$ and $\lambda_R$ is the Rashba spin-orbit coupling strength. It captures spin–momentum locked states and preserves both time reversal ($\Theta$) and the four-fold rotational ($C_{4v}$) symmetries. All energy scales, including hopping amplitudes and interaction strengths, are expressed in the units of $t_J$. We use $\lambda_R/t_J = 1$ throughout the calculations.

Interlayer tunneling between AM and TI includes both spin-conserving and spin-flip processes, described by
\begin{align}
H_{\mathrm{int}} &= \sum_{\sigma,\sigma'} \Big[ 
t_1\, d^\dagger_{\mathbf{k},\sigma} c_{\mathbf{k},\sigma} 
+ t_2\, d^\dagger_{\mathbf{k},\sigma} c_{\mathbf{k},\sigma'} 
+ \text{H.c.} \Big], \label{H_int}
\end{align}

where $t_1 ~\text{and}~ t_2$ denote the spin-conserving and spin-flip tunneling amplitudes, respectively. In a more realistic model, one can explicitly include momentum-dependent tunneling between the two layers (see, for example, \cite{David_PRB2019}). In the presence of an external magnetic field, the total Hamiltonian in Eq.~\ref{H_total} acquires an additional Zeeman contribution
\begin{align}
H_z = \frac{1}{2} g\mu_B \sum_{\sigma,\sigma'}
\Big[ d^{\dagger}_{\mathbf{k},\sigma}
(\mathbf{B}\!\cdot\!\boldsymbol{\sigma})_{\sigma\sigma'} d_{\mathbf{k},\sigma'} + c^{\dagger}_{\mathbf{k},\sigma} (\mathbf{B}\!\cdot\!\boldsymbol{\sigma})_{\sigma\sigma'} c_{\mathbf{k},\sigma'} \Big].
\label{H_z}
\end{align}

where $g$ is the Landé g-factor, $\mu_B$ is the Bohr magneton, and $\textbf{B} = (B_x, \,B_y, \,B_z)$ denotes the external magnetic field vector.

In the absence of tunneling ($t_1=t_2=0$) the bilayer retains the symmetries of the separate layers and can preserve the combined $C_{4z}\Theta$ symmetry if the AM and TI lattices are properly aligned. A finite spin-conserving coupling ($t_1\neq0$, $t_2=0$) generally preserves these symmetries as long as the tunneling amplitude is uniform and identical on both sublattices (momentum-independent, sublattice-symmetric $t_1$). In contrast, spin-flip tunneling ($t_2\neq0$) breaks the remaining spin and spatial symmetries (i.e., it removes $C_{2z}$, $M_x$, $M_y$, and the combined $C_{4z}\Theta$). However, when the two tunneling channels are equal ($t_1=t_2=t$), the mirror symmetry $M_x$ is preserved, protecting a gapless mode along the mirror line.

To analyze spin structure we compute eigenstates $|\psi_{n\mathbf{k}}\rangle$ of $H(\mathbf{k})$ and evaluate the spin expectation value
\begin{align}
\mathbf{S}^n(\mathbf{k}) &= \langle \psi_{n\mathbf{k}} | \boldsymbol{\sigma} | \psi_{n\mathbf{k}} \rangle, \label{SpinTexture}
\end{align}
which yields spin textures across the Brillouin zone.

Transport properties are obtained within linear-response theory. The transverse conductivity is given by the Kubo formula
\begin{equation}
    \sigma_{xy}^{\text{AHE}} = \frac{e^2}{h} \int_{\rm BZ} \frac{d^2\mathbf{k}}{(2\pi)^2}  \sum_n \Omega^n_z(\mathbf{k}) f(\epsilon_n), \label{kubo}
\end{equation}
where the Berry curvature for the n$^{th}$ band is given by
\begin{equation}
    \Omega_z^n(\mathbf{k}) = -2 ~\text{Im} \sum_{m \ne n} \frac{\langle u_n (\mathbf{k})|v_x|u_m (\mathbf{k}) \rangle \langle u_m(\mathbf{k}) |v_y |u_n (\mathbf{k}) \rangle}{(\epsilon_n - \epsilon_m)^2}, \label{berry_curv}
\end{equation}

$v_{x,y} = \partial H / \partial k_{x,y}$, $\epsilon_n$ and $|u_n(\mathbf{k})\rangle$ are the band energies and the eigenstate of the $n^{\text{th}}$ band, and $f(\epsilon_n)$ is the Fermi function \cite{Xiao_RMP2010,Mahan_ManyParticle}.

The current-induced spin polarization at the interface is quantified through the Rashba–Edelstein tensor $\boldsymbol{\chi}$, defined as
\begin{equation}
\label{eq:RE_definition}
S_i= \chi_{ij} E_j ,
\end{equation}
where $i,j \in \{x,y\}$. Within the relaxation time ($\tau$) approximation, the expression is given by \cite{Edelstein_SSC1990, Rojas_NatCommun2013, Manchon_RMP2019}
\begin{equation}
\chi_{ij}
=e\tau \sum_n \int_{\rm BZ} \frac{d^2k}{(2\pi)^2}~
v^n_{j}(\mathbf k)\, S^n_{i}(\mathbf k)\,
\Big(-\frac{\partial f}{\partial \varepsilon_{n}}\Big),
\label{Edelstein}
\end{equation}
Using this framework, both intrinsic anomalous Hall and Rashba-Edelstein responses are evaluated, revealing symmetry-driven anisotropies controlled by interlayer tunneling.

We also construct minimal model Hamiltonians for FM–TI and AFM–TI bilayers. The FM layer is modeled as an itinerant ferromagnet with magnetization along the $z$ direction, while the AFM layer consists of two magnetic sublattices with antiparallel spins characterized by a N\'eel vector along $z$. 
In the basis $(f_{\mathbf{k}\uparrow},\, f_{\mathbf{k}\downarrow})^{T}$ for the FM and $(a_{\mathbf{k}A,\uparrow},\, a_{\mathbf{k}A,\downarrow},\, a_{\mathbf{k}B,\uparrow},\, a_{\mathbf{k}B,\downarrow})^{T}$ for the AFM, where $f_{\textbf{k}\sigma}$ and $a_{\textbf{k} \alpha \sigma}$ annihilate electron in the respective magnetic layers,  with $\alpha \in \{A, B\}$. The corresponding Hamiltonians are written as
\begin{align}
H_{\mathrm{FM}}(\mathbf{k}) &= \epsilon_{\mathbf{k}}\, \sigma_{0} + M \, \sigma_{z} , \nonumber \\
H_{\mathrm{AFM}}(\mathbf{k}) &= \epsilon_{\mathbf{k}}\,\tau_{0}\sigma_{0} + M \, \tau_{z}\sigma_{z} + \Delta \, \tau_0 \, \sigma_x,
\label{HFM_AFM}
\end{align}
where $\epsilon_{\mathbf{k}}$ denotes the band dispersion of the magnetic layers, $M$ is the exchange strength in both FM and AFM layers, and $\Delta$ is the hybridization between the sublattices. 
These minimal models capture the essential magnetic order while neglecting details such as spin-orbit coupling in the magnetic layers, as the dominant spin-orbit effects originate from the TI surface. When coupled to the TI through tunneling processes, these Hamiltonians are used for evaluating the Rashba-Edelstein tensor $\chi_{ij}$.

\begin{figure*}[t]
\begin{center}
\vspace{-0mm}
\epsfig{file=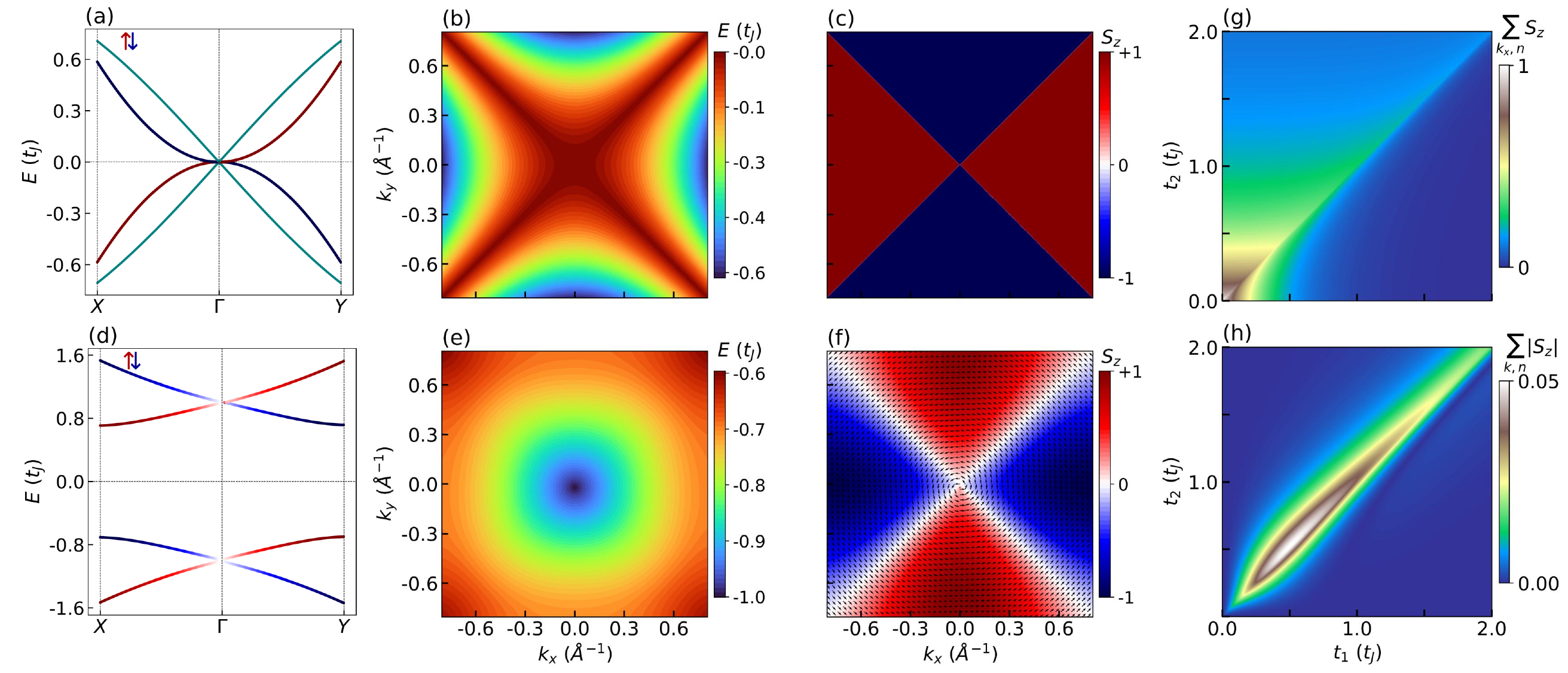,trim=0.0in 0.0in 0.0in 0.0in,clip=false, width=176mm}
\caption{
Proximity-induced altermagnetic features in Rashba–topological-insulator bands. 
Plots (a–c) show the decoupled limit ($t_1=t_2=0$): 
(a) band structure with spin-resolved altermagnet bands and Dirac states (teal) carrying zero out-of-plane spin polarization, 
(b) energy eigenvalues of the lower AM band exhibiting zero-energy modes protected along mirror lines, 
and (c) the corresponding spin polarization. 
Plots (d–f) illustrate the coupled case ($t_1=1.0, t_2=0.01~ t_J$): 
(d) hybridized bands where the lower valence and upper conduction bands acquire finite out-of-plane spin polarization, 
(e) The band dispersion corresponding to the upper valence branch evolves into a Dirac-like form with circular symmetry near the $\Gamma$ point, 
and (f) an inverse $d$-wave spin polarization pattern also emerges, 
accompanied by the in-plane texture reflecting the point of degeneracy with the lower valence branch.
Plot (g) shows the net spin polarization of all occupied states along $k_x$ ($k_y=0$), 
which approaches zero for $t_2<t_1$, while the total $k$-space sum vanishes in both $t_1<t_2$ and $t_1>t_2$ limits depicted in (h).
}
\label{fig2}
\vspace{-4mm}
\end{center}
\end{figure*}

\begin{figure*}[t]
\begin{center}
\vspace{-0mm}
\epsfig{file=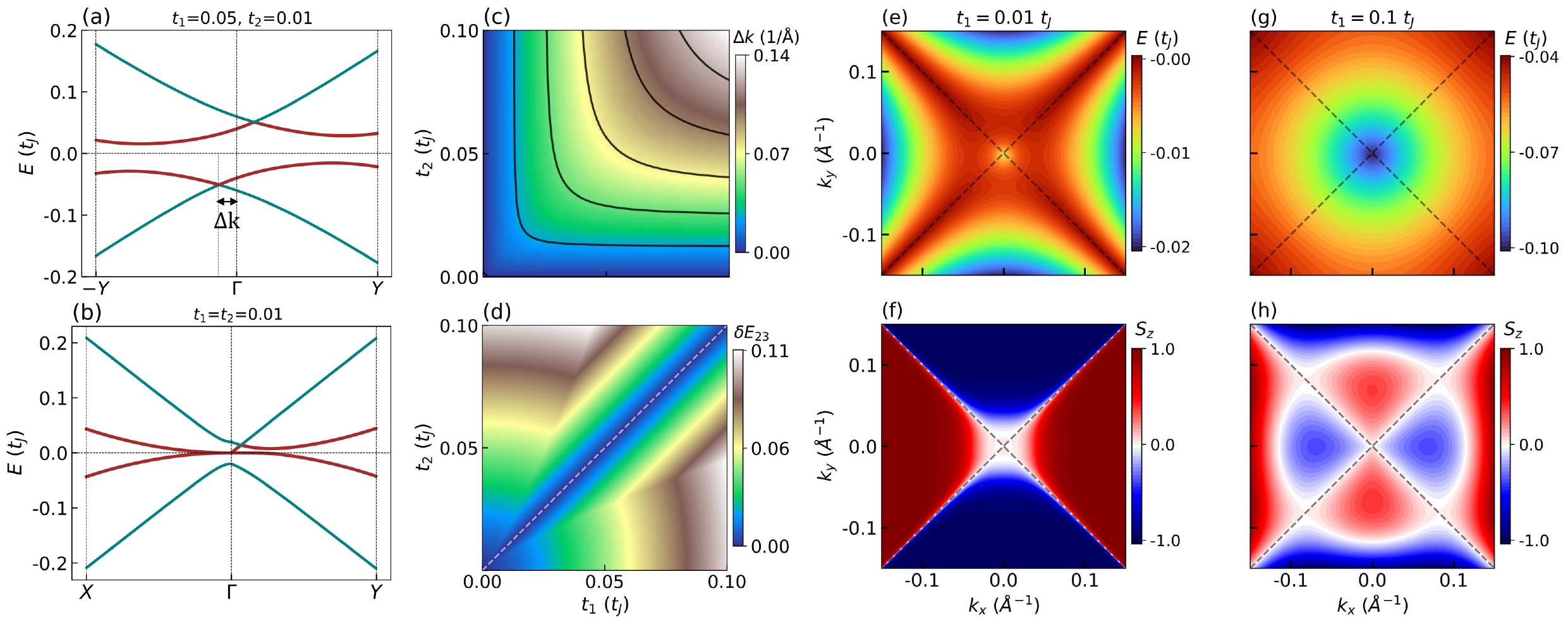,trim=0.0in 0.0in 0.0in 0.0in,clip=false, width=176mm}
\caption{Evolution of eigenvalue spectrum and spin polarization with tunneling amplitudes. (a) Band dispersion along $-\mathrm{Y}\!-\!\Gamma\!-\!\mathrm{Y}$ for $t_{1}=0.05~t_J$, $t_{2}=0.01~t_J$; the degeneracy of occupied (unoccupied) bands shifts by $\Delta k$ along $-k_y$ ($+k_y$). (b) Dispersion along $\mathrm{X}\!-\!\Gamma\!-\!\mathrm{Y}$ for symmetric tunneling $t_{1}=t_{2}=0.01~ t_J$, where a protected zero mode remains along mirror lines. (c) Colormap of $\Delta k$ in the $(t_{1},t_{2})$ plane with analytical contours from Eq.~\eqref{DeltaR}. (d) Minimum inter-band splitting $\delta E_{23}$ showing gap closure and reopening along $t_1=t_2$. (e,f) Energy dispersion and spin polarization of the occupied AM band for $t_2/t_1=0.1$. (g,h) Corresponding results for $t_2/t_1=0.01$, where the texture becomes nearly isotropic.}
\label{fig3}
\vspace{-4mm}
\end{center}
\end{figure*}

\section{Results}

\subsection{Proximity-induced altermagnetism in topological insulators}
\label{proximity_AM_TI}

We first examine how proximity coupling between a $d$-wave AM and a TI reshapes the electronic spectrum and spin textures. The interlayer tunneling consists of a spin-conserving channel $t_{1}$ and a spin-flip channel $t_{2}$, whose relative strength governs the evolution from a decoupled to a hybridized altermagnetic topological phase. 

In the decoupled limit (\(t_1=t_2=0\)), the two subsystems (AM and TI) retain their intrinsic electronic structures, as illustrated in Fig.~\ref{fig2}(a)–(c). The overall band dispersion in Fig.~\ref{fig2}(a) shows spin-resolved AM states along with the teal-colored TI surface states. The TI states exhibit a gapless Dirac cone with a Kramers-degenerate crossing at the \(\Gamma\) point. A four-fold degeneracy arises at the \(\Gamma\) point due to the overlap of the AM and TI states. The TI surface states possess a helical in-plane spin texture originating from spin–momentum locking, while their out-of-plane component \(S_z\) is exactly zero owing to the gapless Dirac nature of the states. The AM energy eigenvalues, shown in Fig.~\ref{fig2}(b), host zero-energy eigenvalues protected along the mirror-symmetric lines \(k_x=\pm k_y\), producing line degeneracies that are characteristic of \(d\)-wave altermagnetism. The corresponding spin texture in Fig.~\ref{fig2}(c) exhibits alternating \(S_z\) lobes with four-fold symmetry and vanishing in-plane components, consistent with the underlying crystal symmetry of the altermagnet.

For finite interfacial coupling (\(t_1 \neq 0\), \(t_2 = 0\)), the AM–TI bilayer undergoes notable band reconstruction. The mirror-protected degeneracies of the lower conduction and upper valence bands are  lifted, signifying the breaking of mirror symmetry at the interface. The proximity effect transfers the characteristic altermagnetic spin texture to the upper conduction and lower valence bands, leading to hybrid spin configurations with both in-plane and out-of-plane components. A finite gap develops at the \(\Gamma\) point, where the two valence bands hybridize below the Fermi level, to form a degenerate state. This hybridized state gradually shifts away from the Fermi level as \(t_1\) increases, with a similar evolution observed for the conduction bands as well. Despite the gap opening and band hybridization, the degeneracy point at the \(\Gamma\) point preserves the helical in-plane spin polarization characteristic of the TI surface states, reflecting the robustness of the topological spin structure under interfacial coupling.

For finite \(t_1\) and \(t_2\) with \(t_1 > t_2\), the AM–TI bilayer undergoes a reconstruction of its electronic and spin structure. The degeneracy point between the conduction bands and valence bands shifts along \(\pm k_y\), with the displacement scaling as \(\Delta k \propto t_1 t_2\) (see Sec. \ref{analytical} for details). The proximity coupling transfers the spin polarization of the upper valence and lower conduction bands to the lower valence and upper conduction bands, while the former develops an inverse \(d\)-wave spin texture for finite \(t_2\). For a fixed \(t_2\), this inverse \(d\)-wave pattern expands progressively across the Brillouin zone as \(t_1\) increases, leading to the emergence of an altermagnetic topological-insulator phase [Fig.~\ref{fig2}(d)]. The evolution of eigenvalues and spin textures in the upper valence band for a fixed \(t_2\) is shown in Figs.~\ref{fig3}(e)–(h), illustrating the gradual growth of the inverse pattern and the lifting of mirror-protected degeneracies. The eigenvalues corresponding to the middle bands near the Fermi level become circularly symmetric with increasing \(t_1\) [Fig.~\ref{fig2}(e)], while the corresponding spin texture [Fig.~\ref{fig2}(f)] displays an intertwined inverse \(d\)-wave and helical in-plane polarization. For \(t_2 > t_1\), the chosen parameters do not support the formation of the inverse \(d\)-wave texture, and the spin configuration remains nearly compensated. The total spin polarization in momentum space remains globally zero regardless of the \(t_1, t_2\) ratio [Fig.~\ref{fig2}(h)], whereas the averaged polarization along \(k_x\) (\(k_y\)) clearly distinguishes the two regimes: it approaches zero for \(t_1 > t_2\) and remains finite for \(t_2 > t_1\), as shown in Fig.~\ref{fig2}(g), marking the emergence of the altermagnetic topological-insulator phase in the former case.

In summary, the interplay between the spin-conserving (\(t_1\)) and spin-flip (\(t_2\)) tunneling channels governs the emergence of altermagnetic topology at the AM–TI interface. For \(t_1 > t_2\), the cooperative action of both channels reconstructs the Dirac and AM bands, transfers the spin texture across the interface, and establishes an inverse \(d\)-wave spin polarization that spreads throughout the Brillouin zone. This regime realizes an altermagnetic topological-insulator phase characterized by circularly symmetric eigenvalues and intertwined helical–inverse \(d\)-wave spin textures. In contrast, when \(t_2 > t_1\), the interface fails to sustain this configuration, resulting in compensated spin textures and the absence of altermagnetic order. Calculations performed using slab geometry as described in Appendix~\ref{appendix:slab_model} show the same interfacial effects, further confirming the robustness of the observed phenomena. The distinct evolution of spin polarization and band symmetry across the two coupling regimes highlights the tunable nature of the AM–TI heterostructure and establishes interfacial tunneling as a key control parameter for engineering spin-compensated topological states.

\subsection{Analytical structure of the spectrum}
\label{analytical}

To complement the numerical findings in Sec.~\ref{proximity_AM_TI}, we analyze the low-energy spectrum of the AM--TI bilayer near the $\Gamma$ point. This analytical treatment clarifies how interfacial tunneling reorganizes band degeneracies and introduces momentum-space displacements that later manifest as transport anisotropies.  

Assuming the N\'eel vector oriented along $\hat{z}$, the low-energy physics is described by the $4\times4$ Bloch Hamiltonian  
\begin{equation}
H(k_x,k_y,t_1,t_2)=
\begin{pmatrix}
A(\bm{k}) & 0 & -t_1 & -t_2\\
0 & -A(\bm{k}) & -t_2 & -t_1\\
-t_1 & -t_2 & 0 & -s_+\\
-t_2 & -t_1 & -s_- & 0
\end{pmatrix},
\label{H4x4_A0}
\end{equation}
where $A(\bm{k})=2(\cos k_x-\cos k_y)$, $s_{\pm}=s_x\pm i s_y$, $s_x=\sin k_x$, and $s_y=\sin k_y$.  
Since $\mathrm{Tr}(H)=0$, the cubic term in the characteristic polynomial vanishes, yielding
\begin{equation}
E^4-\alpha E^2+\beta E+\gamma=0,
\label{char_poly}
\end{equation}
with coefficients
\begin{align}
\alpha &= A^2+s_x^2+s_y^2+2(t_1^2+t_2^2), \nonumber\\
\beta  &= 4\,t_1 t_2\,s_y, \nonumber\\
\gamma &= A^2(s_x^2+s_y^2)+(t_1^2-t_2^2)^2.
\label{char_coeffs}
\end{align}

\textit{One zero-energy mode:} When $\gamma=0$, the spectrum must contain a zero-energy root. This condition is satisfied when both terms in Eq.~\eqref{char_coeffs} vanish: the first term vanishes either along the mirror-symmetric lines $k_x=\pm k_y$ ($A=0$) or at the $\Gamma$ point, while the second requires $t_1=t_2$ or $t_1=t_2=0$. A nontrivial case arises for symmetric hopping $t_1=t_2=t$ and $A=0$, where Eq.~\eqref{char_poly} simplifies to  
\begin{equation}
E^4 - (s_x^2 + s_y^2 + 4t^2)E^2 + 4t^2 s_y E = 0,
\label{gamma=0}
\end{equation}
guaranteeing a single zero-energy mode. This establishes that symmetric interfacial tunneling protects a single zero mode along the mirror lines—unlike the twofold degeneracy of an isolated $d$-wave altermagnet. For $t_1=t_2$ but $A\neq 0$, i.e., slightly off the mirror lines, the resulting dispersion matches the evolution in Fig.~\ref{fig3}(b).  

\textit{Spectrum near the $\Gamma$ point:} Expanding near the $\Gamma$ point yields $\alpha \simeq 4t^2+k_x^2+k_y^2$, $\beta\simeq 4t^2k_y$, and $\gamma\simeq 0$. The vanishing of $\gamma$ signifies a four-fold contact in the decoupled limit, lifted once interfacial tunneling becomes finite. For $k_y=0$, the polynomial reduces to $E^4-(4t^2+k_x^2)E^2=0$, leading to  
\[
E=\{0,\,0,\,\pm\sqrt{4t^2+k_x^2}\},
\]
i.e., two flat zero modes and two gapped branches $\pm 2t$ with quadratic dispersion.  
For $k_x=0$, one obtains  
\begin{equation}
E(E-k_y)(E^2+k_yE-4t^2k_y)=0,
\end{equation}
producing two low-energy solutions $E=0$ and $E=k_y$, along with gapped modes at $E\simeq \pm\sqrt{4t^2+k_y^2}$. The four-fold $\Gamma$-point degeneracy thus reorganizes into one exact zero mode, one linear branch odd in $k_y$, and two gapped states near $\pm 2t$, consistent with Fig.~\ref{fig3}(b). This reorganization underlies the formation of linear modes and their contribution to the Berry curvature near the $\Gamma$ point.

\textit{Two zero-energy modes:} When both $\gamma=0$ and $\beta=0$, two zero-energy roots emerge. At the $\Gamma$ point and for symmetric hopping, $\alpha\simeq 4t^2$, $\beta=0$, and $\gamma\simeq 0$, reducing Eq.~\eqref{char_poly} to $E^4-(4t^2)E^2=0$ with solutions  
\[
E=\{0,\,0,\,\pm 2t\}.
\]
This condition produces the gap closing seen in Fig.~\ref{fig3}(d), corresponding to the vanishing Hall conductivity at $t_1=t_2$ discussed in Sec.~\ref{hall_rashba}. As $t_1$ varies at fixed $t_2$, the minimum interband separation $\delta E_{23}$ remains nearly constant away from the symmetric line, decreases linearly as $t_1\to t_2$, vanishes at $t_1=t_2$, and reopens linearly for $t_1>t_2$.

\textit{Shift of the degeneracy point:} For $t_1=0$ or $t_2=0$, the AM and TI layers decouple and the crossing remains pinned at the $\Gamma$ point. Switching on both couplings shifts the degeneracy along $k_y$, driven by the $\beta$ term, which is odd under $k_y\!\to\!-k_y$ and proportional to $t_1t_2$. Since no analogous term exists in $k_x$, the displacement occurs exclusively along $k_y$. Expanding Eq.~\eqref{char_poly} near the $\Gamma$ point with $k_x=0$ and $|k_y|\ll1$ gives  
\begin{equation}
E^4-2(t_1^2+t_2^2)E^2+4t_1t_2k_yE+(t_1^2-t_2^2)^2=0.
\label{reduced_poly}
\end{equation}
From the two lower branches, the direct energy gap is  
\begin{align}
\delta E(k_y)=k_y+\tfrac12\!\left[\sqrt{k_y^2+4(t_1+t_2)^2}-\sqrt{k_y^2+4(t_1-t_2)^2}\right],
\label{deltaE}
\end{align}
and minimizing $\delta E(k_y)$ yields the band-crossing displacement  
\begin{equation}
\Delta k=\frac{2|t_1t_2|}{\sqrt{t_1^2+t_2^2}}, \qquad |k_y|<1,
\label{DeltaR}
\end{equation}
in lattice units ($a=1$). The conduction and valence crossings shift oppositely along $\pm k_y$, as seen in Fig.~\ref{fig3}(a), while the dependence of $\Delta k$ on $(t_1,t_2)$ in Fig.~\ref{fig3}(c) agrees closely with the analytic contours of Eq.~\eqref{DeltaR}. For symmetric tunneling $t_1=t_2=t$, the result simplifies to $\Delta k=2|t|$, i.e., linear in $t$.

\textit{Summary:} Symmetric interfacial hopping produces a single zero mode along mirror lines, while near the $\Gamma$ point the four-fold degeneracy reorganizes into one exact zero mode, a linear branch odd in $k_y$, and a pair of gapped states. The simultaneous condition $\beta=\gamma=0$ yields two zero modes at the $\Gamma$ point, corresponding to the gap closing at $t_1=t_2$ and reopening for $t_1\neq t_2$. Finite $t_1,t_2$ further shift the crossings strictly along $k_y$ with $\Delta k\propto t_1t_2/\sqrt{t_1^2+t_2^2}$. These analytic results capture the topology of the low-energy manifold and provide benchmarks for Fig.~\ref{fig3}, clarifying the symmetry origins of the transport behavior discussed in Sec.~\ref{hall_rashba}.

\begin{figure}[t]
\begin{center}
\vspace{-0mm}
\epsfig{file=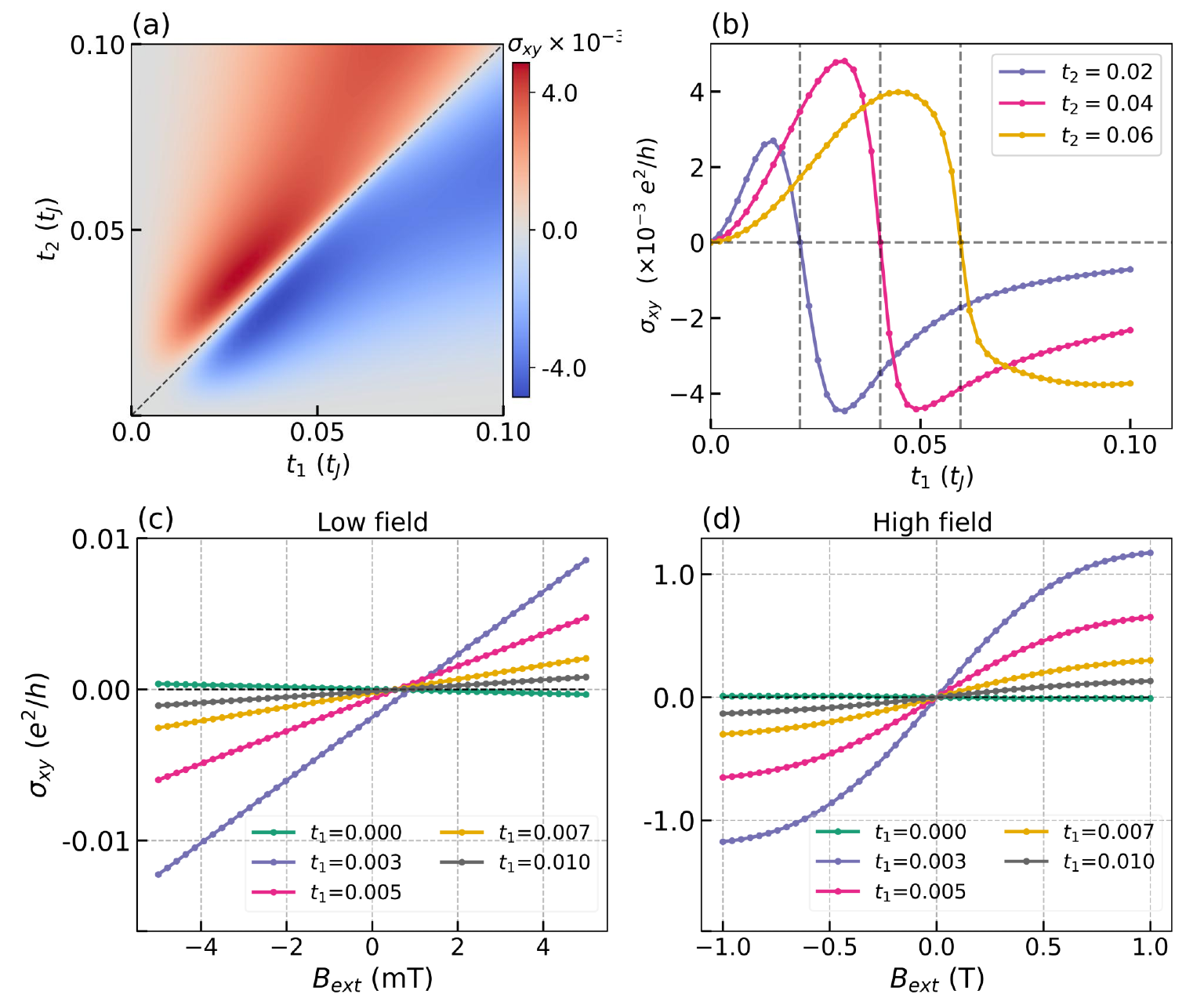,trim=0.0in 0.0in 0.0in 0.0in,clip=false, width=86mm}
\caption{Anomalous Hall conductivity of the AM–TI bilayer. (a) Colormap of $\sigma_{xy}$ in the $(t_{1},t_{2})$ plane, showing a sign reversal across $t_{1}\!\approx\! t_{2}$. (b) $\sigma_{xy}(t_{1})$ at fixed $t_{2}=\{0.02,0.04,0.06\}~t_J$, exhibiting a non-monotonic evolution and zero crossing near $t_{1}\!\sim\!t_{2}$. (c) Low-field $\sigma_{xy}(B_{\mathrm{ext}})$ illustrating sensitivity to weak perturbations. (d) High-field $\sigma_{xy}(B_{\mathrm{ext}})$ showing recovery of antisymmetric behavior. The temperature in the Fermi function is set to $30$ K with hopping parameters measured in $t_J$. For plots (c) and (d), the spin flip tunneling is $0.001~t_J$.}
\label{fig4}
\vspace{-4mm}
\end{center}
\end{figure}

\begin{figure}[t]
\begin{center}
\vspace{-0mm}
\epsfig{file=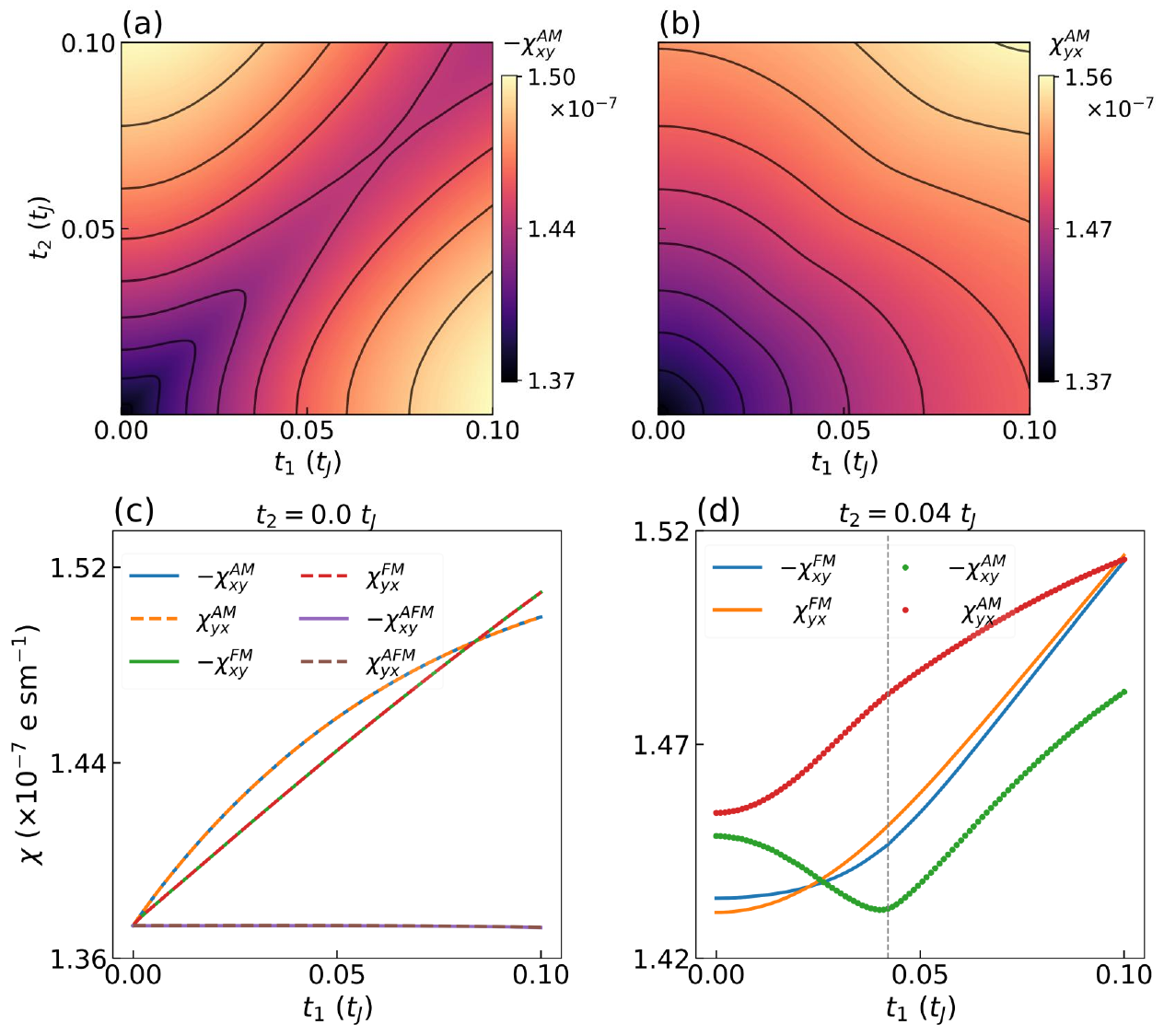,trim=0.0in 0.0in 0.0in 0.0in,clip=false, width=86mm}
\caption{Rashba–Edelstein response tensor of the AM–TI bilayer. (a,b) Colormaps of $\chi_{xy}$ and $\chi_{yx}$ in the $(t_{1},t_{2})$ plane, expressed in $e\,s\,m^{-1}$. Finite $t_{2}$ breaks the antisymmetric relation $\chi_{xy}=-\chi_{yx}$, producing in-plane anisotropy. (c,d) Line cuts of $\chi_{xy}$ and $\chi_{yx}$ comparing FM–TI, AM–TI, and AFM–TI bilayers for $t_{2}=0.0$ and $0.04~t_J$, showing that the AM–TI interface exhibits the strongest deviation from antisymmetry. Other parameters are $T=30$ K, $\tau = 10^{-13}$ s, $M/t_J=1$ and, $\Delta=0.1~t_j$}.
\label{fig5}
\vspace{-4mm}
\end{center}
\end{figure}

\subsection{Hall conductivity and Rashba--Edelstein responses}
\label{hall_rashba}

The interplay between the spin-conserving ($t_{1}$) and spin-flip ($t_{2}$) tunneling channels not only reorganizes the electronic structure but also governs the Berry-curvature–driven spin–charge responses of the AM--TI bilayer. We evaluate both the anomalous Hall conductivity $\sigma_{xy}$ and the Rashba--Edelstein tensor $\boldsymbol{\chi}$, whose evolution directly reflects the hybridized spin textures described in the previous sections. Table~\ref{table1} summarizes the interpretation of the calculated responses reported below.

\begin{table*}[t]
    \centering
    \renewcommand{\arraystretch}{2.2}
    \caption{Symmetries and transport outcomes for the AM–TI bilayer with N\'eel vector along $z$ ($\mathbf n \parallel \hat{z}$).}
    \label{table1}
    \small
    \begin{tabular}{c c c c c c c} \hline
        Configuration & \makecell{~~~Symmetry~~~} & \makecell{States near the $\Gamma$ point} & $~~~~\sigma_{xy}$ ($B=0$)~~~~ & $\Delta k$ & \makecell{~~~~~~~~$\sigma_{xy}(B\neq0)$~~~~~~~~} & ~Edelstein tensor~ \\ \hline
        
        TI layer only & $C_{4v}$, $\Theta$ & Gapless Dirac & Zero & Zero & \makecell{$\sigma_{xy}(B)=-\sigma_{xy}(-B)$ \\ (antisymmetric)} & \makecell{$\chi_{xy}=-\chi_{yx}$ \\ (antisymmetric)} \\
        
        \makecell{AM only \\($\mathbf n\parallel \hat{z}$)} & \makecell{$C_{4z}\Theta$, $C_{2z}$ \\($\Theta$ broken)} & Spin-degenerate & Zero & Zero & $\sigma_{xy}(B)=-\sigma_{xy}(-B)$ & No SOC \\
        
        \makecell{AM$~\oplus~$TI \\($t_1=t_2=0$)} & \makecell{$C_{4z}\Theta^{*}$, $C_{2z}$} & \makecell{Decoupled, \\gapless TI state} & Zero & Zero & $\sigma_{xy}(B)=-\sigma_{xy}(-B)$ & $\chi_{xy}=-\chi_{yx}$ \\
        
        \makecell{AM$~\oplus~$TI, \\($t_1\neq0$, $t_2=0$)} & \makecell{$C_{4z}\Theta^{\dagger}$, $C_{2z}$} & Gapped & Finite & Zero & $\sigma_{xy}(B)=-\sigma_{xy}(-B)$ & $\chi_{xy}=-\chi_{yx}$ \\
        
        \makecell{AM$~\oplus~$TI, \\($t_1=0$, $t_2\neq0$)} & None & Gapped & Finite & Zero & $\sigma_{xy}(B)=-\sigma_{xy}(-B)$ & $\chi_{xy}\neq-\chi_{yx}$ \\
        
        \makecell{AM$~\oplus~$TI, \\($t_1=t_2=t$)} & $M_{x}$ & \makecell{Gapless, one \\zero mode along \\the mirror line } & Zero & $2|t|$ & $\sigma_{xy}(B)=-\sigma_{xy}(-B)$ & $\chi_{xy}\neq-\chi_{yx}$ \\
        
        \makecell{AM$~\oplus~$TI, \\($t_1\neq0$, $t_2\neq0$)} & None & Gapped & \makecell{Finite, sign \\reversal across \\$t_1=t_2$ (vanishes \\at $t_1=t_2$)} & $\dfrac{2|t_1t_2|}{\sqrt{t_1^2+t_2^2}}$ & Non-antisymmetric & $\chi_{xy}\neq-\chi_{yx}$ \\ \hline
    \end{tabular}
    \vspace{4pt}
    \raggedright
    
    {\footnotesize
    $^{*}$Valid only when the AM and TI lattices are perfectly aligned under four-fold rotation. \\[1pt]
    $^{\dagger}$Preserved if the spin-conserving tunneling $t_1$ is momentum-independent and identical on both sublattices.}
\end{table*}

\paragraph{Hall conductivity and gap topology} In the absence of an external field, $\sigma_{xy}$ originates from the Berry curvature generated by interfacial hybridization, which simultaneously breaks time-reversal and inversion symmetries. As shown in Fig.~\ref{fig4}(a), the sign of $\sigma_{xy}$ depends critically on the relative strengths of the tunneling channels: it is negative for $t_2 < t_1$, positive for $t_2 > t_1$, and vanishes at $t_1 = t_2$, where the hybridization gap closes. The magnitude of $\sigma_{xy}$ exhibits a nonmonotonic evolution with increasing coupling, reaching a maximum at intermediate strengths and decreasing thereafter [Fig.~\ref{fig4}(b)]. This behavior stems from the dependence of the Berry curvature on the hybridization gap. Initially, an increasing gap enhances the curvature and thus $\sigma_{xy}$, but for larger coupling, the curvature becomes more localized in momentum space, leading to its suppression. The displacement of the degeneracy point, $\Delta k \!\propto\! t_1t_2/\sqrt{t_1^2+t_2^2}$, further contributes to this trend by shifting the Berry-curvature center away from the $\Gamma$ point. Including a Rashba mass term preserves a finite gap throughout and removes the sign reversal.

Under an external field \(B_{\mathrm{ext}}\) the symmetry of the Hall response changes qualitatively. For single-channel coupling (\(t_2=0\)) the Hall conductivity remains antisymmetric under field reversal, \(\sigma_{xy}(B_{\mathrm{ext}}) = -\sigma_{xy}(-B_{\mathrm{ext}})\), in line with conventional magnetotransport. When both \(t_1\) and \(t_2\) are finite, however, \(\sigma_{xy}\) develops anisotropy and loses this antisymmetry in the intermediate-field regime where interfacial hopping and Zeeman energy scales compete, as seen in Fig.~\ref{fig4}(c). This unconventional behavior arises from the combined effect of proximity-induced Berry curvature and field-driven band reconstruction, which modify the Berry-curvature distribution asymmetrically under opposite field orientations. In the high-field limit \(B_{\mathrm{ext}}\gg t_{1,2}\) the Zeeman term dominates and the Hall response recovers the expected antisymmetric form shown in Fig.~\ref{fig4}(d).

\paragraph{Rashba--Edelstein response and anisotropy}
The current-induced spin polarization, described by the tensor $\boldsymbol{\chi}$ (Eq.~\ref{Edelstein}), reflects the sensitivity of the interfacial spin texture to tunneling between the AM and TI layers. In Figs.~\ref{fig5}(a) and \ref{fig5}(b), the overall magnitude $|\boldsymbol{\chi}|$ increases steadily with either spin-conserving ($t_1$) or spin-flip ($t_2$) tunneling, due to enhanced interlayer hybridization. By anisotropy we mean the deviation from the conventional antisymmetric relation $\chi_{xy}=-\chi_{yx}$. Finite $t_{2}$ breaks the in plane mirror symmetry and produces this deviation.  The dependence of $\boldsymbol{\chi}$ on $(t_1,t_2)$ is not purely monotonic. The component $\chi_{xy}$ shows a clear dip along the symmetric line $t_1=t_2$ in Fig.~\ref{fig5}(a) and \ref{fig5}(d), marking a regime where the hybridization gap closes and the Berry-curvature distribution is reorganized. Away from that line both off diagonal components vary smoothly and the magnitude of the deviation grows approximately with $t_{2}$.

Figures~\ref{fig5}(c) and \ref{fig5}(d) compare the AM--TI results with FM--TI and AFM--TI modeled using the Hamiltonian given in Eq. \ref{HFM_AFM}. For $t_{2}=0$ all systems satisfy $\chi_{xy}=-\chi_{yx}$. For $t_{2}\neq0$ the breakdown of this antisymmetry depends on the magnetic order. The deviation is weakest in FM--TI, moderate in AFM--TI, and strongest in AM--TI where spin flip tunneling is most effective. FM--TI exhibits the largest overall enhancement of $|\boldsymbol{\chi}|$ with both off diagonal components increasing substantially. AFM--TI shows negligible enhancement of $|\boldsymbol{\chi}|$ because of partial sublattice cancellation. In AM--TI the $\chi_{yx}$ component grows strongly and in large regions of the $(t_{1},t_{2})$ plane its magnitude exceeds the overall enhancement of $|\boldsymbol{\chi}|$ seen in FM--TI, producing the dominant nonantisymmetric response. Thus the AM--TI interface combines the large magnitude enhancement characteristic of FM--TI with the pronounced deviation from antisymmetry characteristic of AFM--TI while remaining free of a net stray field.

\paragraph{Correlation between band topology and spin-charge responses} Both the anomalous Hall conductivity $\sigma_{xy}$ and the Rashba--Edelstein tensor $\boldsymbol{\chi}$ trace back to the same microscopic origin. Interfacial tunneling reshapes the low-energy band manifold and displaces the Berry-curvature hotspot in momentum space by $\Delta k$, which in turn alters the integrated Berry flux that determines $\sigma_{xy}$ and redistributes the momentum-resolved spin accumulation that enters $\boldsymbol{\chi}$. A detailed description of the role of Berry curvature is given in Appendix~\ref{appendix:berry_curvature}. The combined effect produces the observed sign reversal of $\sigma_{xy}$ and the breakdown of the Edelstein antisymmetry relation in $\boldsymbol{\chi}$. In this picture the spin-flip amplitude $t_2$ provides a direct tuning knob that links band topology to spin–charge conversion. Together these findings position AM--TI heterostructures as a controllable, stray-field free platform for direction-sensitive spin–charge interconversion driven by proximity-induced altermagnetic correlations.

\vspace{0.2cm}

\section*{Discussion}

Our results show that interfacial spin-flip tunneling at $d$-wave AM–TI interfaces produces pronounced band reconstructions and hybrid spin textures that are qualitatively distinct from those in ferromagnet–TI or antiferromagnet–TI heterostructures. In particular, the induced inverse $d$-wave $S_z$ pattern—together with retained TI-like in-plane winding—serves as a characteristic signature of an altermagnetic topological insulating phase. These texture changes are accompanied by strong anisotropies in both Rashba–Edelstein and Hall responses, demonstrating that symmetry and interlayer spin-mixing together control direction-sensitive spin–charge interconversion without relying on stray magnetic fields. The relative strength of these anisotropies is consistent with the symmetry constraints listed in Table \ref{table1}.

From a materials perspective, several realistic platforms can be used to implement the bilayers studied here. On the altermagnet side, RuO$_2$ has been shown by photoemission to host band spin splitting consistent with altermagnetism \cite{Fedchenko_SciAdv2024,Guo_AdvSci2024}, while MnTe exhibits giant anisotropic band splitting confirmed by spectroscopic studies and nanoscale imaging \cite{Osumi_PRB2024,Amin_Nature2024}. CrSb thin films also provide direct experimental evidence of altermagnetic band splitting \cite{Reimers_NatCommun2024}. On the TI side, Bi$_2$Se$_3$ and Bi$_2$Te$_3$ (and related alloys) remain as the benchmark materials with robust Dirac surface states and high-quality thin-film growth \cite{Hasan_RMP2010,Qi_Zhang_RevModPhys2011}. 

Interfacial engineering offers viable routes to tune the relative strength of tunneling channels. Heavy-element oxides such as SrIrO$_3$ have demonstrated strong spin–orbit torque generation \cite{Nan_PNAS2019}, and dichalcogenides like WTe$_2$ provide large intrinsic SOC \cite{Ali_WTe2_Nature2014}, making them attractive as buffer layers to enhance spin-flip tunneling. Conversely, thin insulating spacers such as MgO or Al$_2$O$_3$ can suppress spin-conserving tunneling and allow the spin-flip channel to dominate. These approaches suggest that AM–TI bilayers are experimentally accessible platforms for realizing and tuning symmetry-driven, stray-field–free spin–charge conversion.

\section*{Acknowledgements} 
JS was supported by Ministry of Education, Government of India via a research fellowship. NM acknowledges support of an initiation grant (No. IlTR/SRIC/2116/FIG) from IIT Roorkee and SRG grant (No. SRG/2023/001188) from SERB. 

\appendix

\section{Slab model of the AM--TI heterostructure}
\label{appendix:slab_model}
\noindent
To explicitly capture the layer-resolved coupling between the altermagnet (AM) and the topological insulator (TI), we construct the total slab Hamiltonian for an AM layer stacked on top of $N$ quintuple layers (QLs) of Bi$_2$Se$_3$. The full Hamiltonian in momentum space is expressed as
\begin{equation}
    H(\mathbf{k}) =
    \begin{pmatrix}
        H_{\mathrm{AM}} & H_{\mathrm{int}} & & & & \\
        H^{\dagger}_{\mathrm{int}} & H_0 & H_1 &  &   &   &  &  \\
        & H^{\dagger}_1 & H_0 & H_1 &  &  &  &  \\
        & & H^{\dagger}_1 & H_0 & H_1 &   &  &  \\
        &  & &  . &   &   &  &  \\
        & &  &   &.   &  &  &  \\ 
        & &  &   &   &.  &  &  \\  
        & &  &  &  &   H^{\dagger}_1 & H_0 & H_1\\
        & &  &  & & &   H^{\dagger}_1 & H_0 \\
    \end{pmatrix},
    \label{H_AM_TI_QL}
\end{equation}

\noindent
written in the composite basis 
$\{\ket{d_{\mathbf{k}},\uparrow}$, $\ket{d_{\mathbf{k}},\downarrow}$,  $\ket{P1^+_z, \uparrow, l_z}$, $\ket{P2^-_z, \uparrow, l_z}$,  $\ket{P1^+_z, \downarrow, l_z}$, $\ket{P2^-_z, \downarrow, l_z}\}$,
where $\mathbf{k}_{\parallel} \!=\! (k_x, k_y)$ and $l_z = 1, 2, \dots, N$ labels the QLs of Bi$_2$Se$_3$. 
Here, $H_{\mathrm{AM}}$, $H_{\mathrm{int}}$, $H_0$, and $H_1$ represent the AM Hamiltonian, AM--TI interfacial coupling, intra-QL Hamiltonian, and inter-QL coupling, respectively. 

\noindent
The AM Hamiltonian $H_{\mathrm{AM}}$ follows Eq.~\ref{H_AM}, while the interfacial tunneling term includes both spin-conserving and spin-flip processes, described by
\begin{equation}
    H_{\mathrm{int}} =
    \begin{pmatrix}
        t_1 & t_1 & t_2 & t_2 \\
        t_2 & t_2 & t_1 & t_1 \\
    \end{pmatrix}.
\end{equation}

\noindent
The TI layers are modeled using a $4 \times 4$ effective Hamiltonian for Bi$_2$Se$_3$. The intra-QL and inter-QL coupling matrices are given by
\begin{equation}
H_0 =
\begin{pmatrix}
  \epsilon_0 + M & 0 & 0 & A_0 k_- \\ 
  0 & \epsilon_0 - M & A_0 k_- & 0 \\ 
  0 & A_0 k_+ & \epsilon_0 + M & 0 \\
  A_0 k_+ & 0 & 0 & \epsilon_0 - M
\end{pmatrix},
\label{H0}
\end{equation}

\begin{equation}
H_1 =
\begin{pmatrix}
  -M_1 - C_1 & i B_0/2 & 0 & 0 \\ 
  i B_0/2 & M_1 - C_1 & 0 & 0 \\ 
  0 & 0 & -M_1 - C_1 & -i B_0/2 \\
  0 & 0 & -i B_0/2 & M_1 - C_1
\end{pmatrix}.
\label{H1}
\end{equation}

\noindent
The momentum-dependent terms are defined as
\[
\epsilon_0(\mathbf{k}_{\parallel}) = C_0 + 2C_1 + 2C_2[2 - \cos(k_x a) - \cos(k_y a)],
\]
\[
M_0(\mathbf{k}_{\parallel}) = M_0 + 2M_1 + 2M_2[2 - \cos(k_x a) - \cos(k_y a)],
\]
\[
\mathbf{k}_{\pm} = \sin(k_x a) \pm i \sin(k_y a),
\]
where $a$ is the lattice constant. The parameters used are 
$A_0 = 0.8~\mathrm{eV}$,  
$B_0 = 0.32~\mathrm{eV}$,  
$C_0 = -0.0083~\mathrm{eV}$,  
$C_1 = 0.024~\mathrm{eV}$,  
$C_2 = 1.77~\mathrm{eV}$,  
$M_0 = -0.28~\mathrm{eV}$,  
$M_1 = 0.216~\mathrm{eV}$,  
$M_2 = 2.6~\mathrm{eV}$,  
and $a = 4.14~\mathrm{\AA}$ \cite{Zhang_NatPhys2009, Liu_modelTI_PRB2010, Mohanta_SciRep2017, JSingh_PRB2024}.

\begin{figure}[t]
\begin{center}
\vspace{-1mm}
\epsfig{file=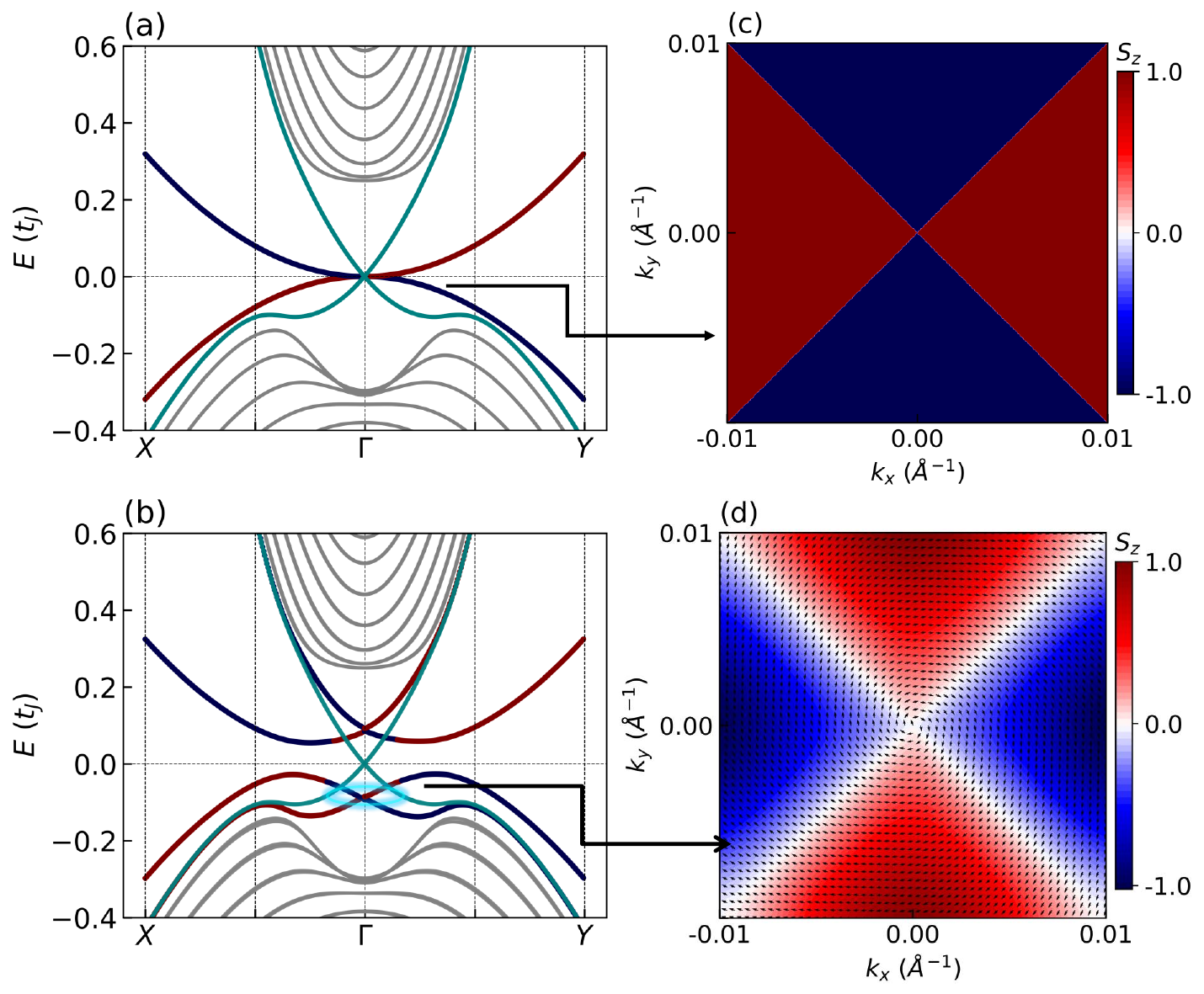,trim=0.0in 0.0in 0.0in 0.0in,clip=false, width=86mm}
\caption{Proximity effect in the AM-TI slab: (a) Band structure in the decoupled limit showing the hybridization between the altermagnetic states and the TI surface Dirac cone. (b) The top-surface Dirac band develops a finite gap at the $\Gamma$ point due to proximity coupling ($t_1=0.1$ and $t_2=0.005~t_J$), while the bottom-surface Dirac cone remains gapless. Teal bands denote the gapless surface states, and grey bands correspond to the bulk states. (c) and (d) depicts the spin textures of the bands marked by arrows in the decoupled and coupled limits, respectively, plotted in the vicinity of the $\Gamma$ point. The number of TI quintuple layers used in all calculations is $N = 10$.}
\label{fig6}
\vspace{-4mm}
\end{center}
\end{figure}

\noindent
The slab geometry calculation provides a microscopic perspective on the minimal model, demonstrating that the essential proximity-induced features are captured even without explicitly including bulk states. In the decoupled limit, the teal-colored bands in Fig.~\ref{fig6}(a) correspond to the Dirac surface states localized at the top and bottom surfaces of the TI. When interfacial coupling is introduced, the top-surface Dirac cone hybridizes with the altermagnet states, opening a gap at the $\Gamma$ point, while the bottom-surface cone remains gapless due to spatial separation, as shown in Fig.~\ref{fig6}(b). The spin texture of the hybridized states in Fig.~\ref{fig6}(c,d) exhibits an inverse $d$-wave spin polarization near the $\Gamma$ point, originating from finite spin-flip tunneling. Meanwhile, the in-plane helical spin texture of the TI surface persists around the degeneracy point [Fig.~\ref{fig6}(d)], confirming that the surface’s helical nature survives under interfacial exchange coupling. These band structure features are consistent with the trends discussed in the main text.

For interfacial tunneling amplitudes $t_1, t_2 < 0.1~ t_J$, the hybridization gap varies as $2|t_1 - t_2|$ at the $\Gamma$ point and reaches its maximum value within the bulk gap, ensuring that only surface states near the Fermi level contribute to the transport phenomena discussed in the main text. The selected range of $t_1$ and $t_2$ falls within the commonly used theoretical parameter window \cite{Gupta_SciRep_2014}. The simplified low-energy model therefore captures the key interfacial physics of the AM–TI heterostructure with quantitative accuracy and conceptual clarity.

\begin{figure}[t]
\begin{center}
\vspace{-1mm}
\epsfig{file=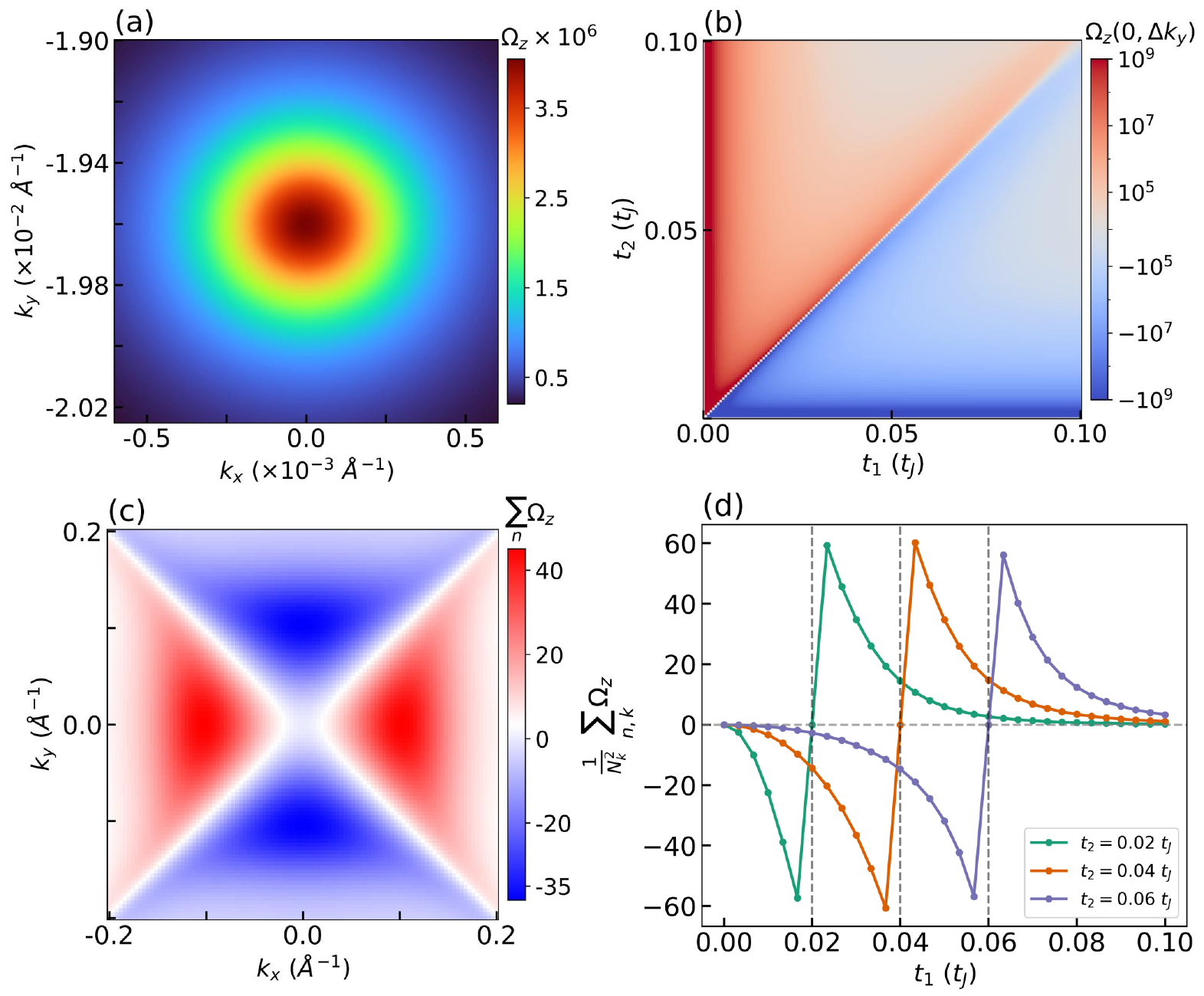,trim=0.0in 0.0in 0.0in 0.0in,clip=false, width=86mm}
\caption{
(a) $k$-resolved Berry curvature for $t_1 = 0.05$ and $t_2 = 0.01~ t_J$. 
(b) Berry curvature at $k = (0, \Delta k)$ for the upper valence band in the $(t_1, t_2)$ parameter space. 
(c) Total Berry curvature from occupied bands below the Fermi level for $t_1 = 0.05$ and $t_2 = 0.01~t_J$. 
(d) Normalized Berry curvature summed over the Brillouin zone as a function of hopping amplitudes. 
The calculations are performed with $N_k = 101$ points along each momentum direction, giving a total of $N_k^2$ $k$-points in the Brillouin zone.}
\label{fig7}
\vspace{-4mm}
\end{center}
\end{figure}

\section{Role of Berry curvature}
\label{appendix:berry_curvature}

To understand the microscopic origin of the anomalous Hall response, we evaluate the Berry curvature for all occupied bands using Eq.~(\ref{berry_curv}). Figure~\ref{fig7}(a) presents the $k$-resolved Berry curvature distribution for $t_1 = 0.05~ t_J$ and $t_2 = 0.01~ t_J$. The curvature exhibits a distinct localization in momentum space, with its maximum shifted from the $\Gamma$ point, coinciding with the displaced degeneracy point of the band structure. The upper and lower valence bands display opposite curvature signs, which partially cancel when their contributions are summed.

Figure~\ref{fig7}(b) shows the Berry curvature at the shifted degeneracy point, $k = (0, \Delta k)$, of the upper valence band in the $(t_1, t_2)$ parameter space. The curvature changes sign across the line $t_1 = t_2$, consistent with the closing and reopening of the band gap at symmetric hopping amplitudes. This behavior reflects a topological transition arising from the hybridization of the altermagnet and topological insulator states. In Fig.~\ref{fig7}(c), the total Berry curvature from all bands below the Fermi level is plotted. Regions with opposite curvature signs cancel, resulting in a pattern resembling the spin-polarization texture. The maxima and minima are asymmetric, and integrating the curvature over the Brillouin zone yields a finite but reduced value due to partial cancellation near the degeneracy point. 

Figure~\ref{fig7}(d) illustrates the normalized Berry curvature summed over the Brillouin zone as a function of the hopping amplitudes. A clear sign reversal occurs as the system crosses the symmetric coupling condition ($t_1 = t_2$), mirroring the variation of the Hall conductivity shown in Fig.~\ref{fig4}(a)–(b). This correspondence indicates that the Berry curvature directly governs the Hall response. The partial cancellation of curvature around the degeneracy point leads to a finite yet non-quantized anomalous Hall effect in the coupled system.

\paragraph*{Spin–charge responses.}
The spin–charge conversion is evaluated through the Rashba–Edelstein response using Eq.~(\ref{Edelstein}), which involves the momentum-dependent velocity $\bm{v}_k$ and spin polarization $\bm{S}$. For $t_2 = 0$, the velocities along the two in-plane directions are equal ($v_x = v_y$), while the spin components exhibit opposite signs ($S_x = -S_y$), generating an  antisymmetric response, $\chi_{xy} = -\chi_{yx}$. When $t_2$ becomes finite, the velocities and spin polarizations evolve anisotropically, breaking this antisymmetry relation. The response function $\chi_{yx}$, which involves the $v_y S_x$ term in the integrand shows an enhancement, while $\chi_{xy}$, governed by $v_x S_y$, decreases for $t_1 < t_2$, attains a minimum near $t_1 = t_2$, and increases again for $t_1 > t_2$. These variations originate from the anisotropic band rearrangements that generate finite Berry curvature in the coupled system. 
The above analysis shows that band rearrangement, the resulting momentum-space Berry curvature, and the anisotropic evolution of velocity and spin textures together determine the observed Hall and Rashba--Edelstein responses in the coupled system.

\bibliography{ref}

\begin{thebibliography}{71}%
\makeatletter
\providecommand \@ifxundefined [1]{%
 \@ifx{#1\undefined}
}%
\providecommand \@ifnum [1]{%
 \ifnum #1\expandafter \@firstoftwo
 \else \expandafter \@secondoftwo
 \fi
}%
\providecommand \@ifx [1]{%
 \ifx #1\expandafter \@firstoftwo
 \else \expandafter \@secondoftwo
 \fi
}%
\providecommand \natexlab [1]{#1}%
\providecommand \enquote  [1]{``#1''}%
\providecommand \bibnamefont  [1]{#1}%
\providecommand \bibfnamefont [1]{#1}%
\providecommand \citenamefont [1]{#1}%
\providecommand \href@noop [0]{\@secondoftwo}%
\providecommand \href [0]{\begingroup \@sanitize@url \@href}%
\providecommand \@href[1]{\@@startlink{#1}\@@href}%
\providecommand \@@href[1]{\endgroup#1\@@endlink}%
\providecommand \@sanitize@url [0]{\catcode `\\12\catcode `\$12\catcode `\&12\catcode `\#12\catcode `\^12\catcode `\_12\catcode `\%12\relax}%
\providecommand \@@startlink[1]{}%
\providecommand \@@endlink[0]{}%
\providecommand \url  [0]{\begingroup\@sanitize@url \@url }%
\providecommand \@url [1]{\endgroup\@href {#1}{\urlprefix }}%
\providecommand \urlprefix  [0]{URL }%
\providecommand \Eprint [0]{\href }%
\providecommand \doibase [0]{http://dx.doi.org/}%
\providecommand \selectlanguage [0]{\@gobble}%
\providecommand \bibinfo  [0]{\@secondoftwo}%
\providecommand \bibfield  [0]{\@secondoftwo}%
\providecommand \translation [1]{[#1]}%
\providecommand \BibitemOpen [0]{}%
\providecommand \bibitemStop [0]{}%
\providecommand \bibitemNoStop [0]{.\EOS\space}%
\providecommand \EOS [0]{\spacefactor3000\relax}%
\providecommand \BibitemShut  [1]{\csname bibitem#1\endcsname}%
\let\auto@bib@innerbib\@empty
\bibitem [{\citenamefont {{Šmejkal}}\ \emph {et~al.}(2022{\natexlab{a}})\citenamefont {{Šmejkal}}, \citenamefont {Sinova},\ and\ \citenamefont {Jungwirth}}]{Smejkal_PRX2022b}%
  \BibitemOpen
  \bibfield  {author} {\bibinfo {author} {\bibfnamefont {L.}~\bibnamefont {{Šmejkal}}}, \bibinfo {author} {\bibfnamefont {J.}~\bibnamefont {Sinova}}, \ and\ \bibinfo {author} {\bibfnamefont {T.}~\bibnamefont {Jungwirth}},\ }\bibfield  {title} {\enquote {\bibinfo {title} {Emerging research landscape of altermagnetism},}\ }\href {\doibase 10.1103/PhysRevX.12.040501} {\bibfield  {journal} {\bibinfo  {journal} {Phys. Rev. X}\ }\textbf {\bibinfo {volume} {12}},\ \bibinfo {pages} {040501} (\bibinfo {year} {2022}{\natexlab{a}})}\BibitemShut {NoStop}%
\bibitem [{\citenamefont {{Šmejkal}}\ \emph {et~al.}(2022{\natexlab{b}})\citenamefont {{Šmejkal}}, \citenamefont {Sinova},\ and\ \citenamefont {Jungwirth}}]{Smejkal_PRX2022a}%
  \BibitemOpen
  \bibfield  {author} {\bibinfo {author} {\bibfnamefont {L.}~\bibnamefont {{Šmejkal}}}, \bibinfo {author} {\bibfnamefont {J.}~\bibnamefont {Sinova}}, \ and\ \bibinfo {author} {\bibfnamefont {T.}~\bibnamefont {Jungwirth}},\ }\bibfield  {title} {\enquote {\bibinfo {title} {Beyond conventional ferromagnetism and antiferromagnetism: A phase with nonrelativistic spin and crystal rotation symmetry},}\ }\href {\doibase 10.1103/PhysRevX.12.031042} {\bibfield  {journal} {\bibinfo  {journal} {Phys. Rev. X}\ }\textbf {\bibinfo {volume} {12}},\ \bibinfo {pages} {031042} (\bibinfo {year} {2022}{\natexlab{b}})}\BibitemShut {NoStop}%
\bibitem [{\citenamefont {{Šmejkal}}\ \emph {et~al.}(2022{\natexlab{c}})\citenamefont {{Šmejkal}}, \citenamefont {MacDonald}, \citenamefont {Sinova}, \citenamefont {Nakatsuji},\ and\ \citenamefont {Jungwirth}}]{Smejkal_NatRevMater2022}%
  \BibitemOpen
  \bibfield  {author} {\bibinfo {author} {\bibfnamefont {L.}~\bibnamefont {{Šmejkal}}}, \bibinfo {author} {\bibfnamefont {A.~H.}\ \bibnamefont {MacDonald}}, \bibinfo {author} {\bibfnamefont {J.}~\bibnamefont {Sinova}}, \bibinfo {author} {\bibfnamefont {S.}~\bibnamefont {Nakatsuji}}, \ and\ \bibinfo {author} {\bibfnamefont {T.}~\bibnamefont {Jungwirth}},\ }\bibfield  {title} {\enquote {\bibinfo {title} {Anomalous {H}all antiferromagnets},}\ }\href {\doibase 10.1038/s41578-022-00430-3} {\bibfield  {journal} {\bibinfo  {journal} {Nat. Rev. Mater}\ }\textbf {\bibinfo {volume} {7}},\ \bibinfo {pages} {482--496} (\bibinfo {year} {2022}{\natexlab{c}})}\BibitemShut {NoStop}%
\bibitem [{\citenamefont {Attias}\ \emph {et~al.}(2024)\citenamefont {Attias}, \citenamefont {Levchenko},\ and\ \citenamefont {Khodas}}]{Attias_PRB2024}%
  \BibitemOpen
  \bibfield  {author} {\bibinfo {author} {\bibfnamefont {L.}~\bibnamefont {Attias}}, \bibinfo {author} {\bibfnamefont {A.}~\bibnamefont {Levchenko}}, \ and\ \bibinfo {author} {\bibfnamefont {M.}~\bibnamefont {Khodas}},\ }\bibfield  {title} {\enquote {\bibinfo {title} {Intrinsic anomalous {H}all effect in altermagnets},}\ }\href {\doibase 10.1103/PhysRevB.110.094425} {\bibfield  {journal} {\bibinfo  {journal} {Phys. Rev. B}\ }\textbf {\bibinfo {volume} {110}},\ \bibinfo {pages} {094425} (\bibinfo {year} {2024})}\BibitemShut {NoStop}%
\bibitem [{\citenamefont {Reichlova}\ \emph {et~al.}(2024)\citenamefont {Reichlova}, \citenamefont {Lopes~Seeger}, \citenamefont {Gonz\'alez-Hern\'andez}, \citenamefont {Kounta}, \citenamefont {Schlitz}, \citenamefont {Kriegner}, \citenamefont {Ritzinger}, \citenamefont {Lammel}, \citenamefont {Leivisk\"a}, \citenamefont {Hellenes}, \citenamefont {Olejn\'ik}, \citenamefont {Pet\u{r}\u{i}\u{c}ek}, \citenamefont {Dole\u{z}al}, \citenamefont {Hor\'ak}, \citenamefont {Schmoranzerov\'a}, \citenamefont {Badura}, \citenamefont {Bertaina}, \citenamefont {Thomas}, \citenamefont {Baltz}, \citenamefont {Michez}, \citenamefont {Sinova}, \citenamefont {Goennenwein}, \citenamefont {Jungwirth},\ and\ \citenamefont {{Šmejkal}}}]{Helena_NatCommun2024}%
  \BibitemOpen
  \bibfield  {author} {\bibinfo {author} {\bibfnamefont {H.}~\bibnamefont {Reichlova}}, \bibinfo {author} {\bibfnamefont {R.}~\bibnamefont {Lopes~Seeger}}, \bibinfo {author} {\bibfnamefont {R.}~\bibnamefont {Gonz\'alez-Hern\'andez}}, \bibinfo {author} {\bibfnamefont {I.}~\bibnamefont {Kounta}}, \bibinfo {author} {\bibfnamefont {R.}~\bibnamefont {Schlitz}}, \bibinfo {author} {\bibfnamefont {D.}~\bibnamefont {Kriegner}}, \bibinfo {author} {\bibfnamefont {P.}~\bibnamefont {Ritzinger}}, \bibinfo {author} {\bibfnamefont {M.}~\bibnamefont {Lammel}}, \bibinfo {author} {\bibfnamefont {M.}~\bibnamefont {Leivisk\"a}}, \bibinfo {author} {\bibfnamefont {A.~B.}\ \bibnamefont {Hellenes}}, \bibinfo {author} {\bibfnamefont {K.}~\bibnamefont {Olejn\'ik}}, \bibinfo {author} {\bibfnamefont {V.}~\bibnamefont {Pet\u{r}\u{i}\u{c}ek}}, \bibinfo {author} {\bibfnamefont {P.}~\bibnamefont {Dole\u{z}al}}, \bibinfo {author} {\bibfnamefont {L.}~\bibnamefont {Hor\'ak}}, \bibinfo {author} {\bibfnamefont {E.}~\bibnamefont
  {Schmoranzerov\'a}}, \bibinfo {author} {\bibfnamefont {A.}~\bibnamefont {Badura}}, \bibinfo {author} {\bibfnamefont {S.}~\bibnamefont {Bertaina}}, \bibinfo {author} {\bibfnamefont {A.}~\bibnamefont {Thomas}}, \bibinfo {author} {\bibfnamefont {V.}~\bibnamefont {Baltz}}, \bibinfo {author} {\bibfnamefont {L.}~\bibnamefont {Michez}}, \bibinfo {author} {\bibfnamefont {J.}~\bibnamefont {Sinova}}, \bibinfo {author} {\bibfnamefont {S.~T.~B.}\ \bibnamefont {Goennenwein}}, \bibinfo {author} {\bibfnamefont {T.}~\bibnamefont {Jungwirth}}, \ and\ \bibinfo {author} {\bibfnamefont {L.}~\bibnamefont {{Šmejkal}}},\ }\bibfield  {title} {\enquote {\bibinfo {title} {Observation of a spontaneous anomalous {H}all response in the {M}n$_5${S}i$_3$ d-wave altermagnet candidate},}\ }\href {\doibase 10.1038/s41467-024-48493-w} {\bibfield  {journal} {\bibinfo  {journal} {Nat. Commun.}\ }\textbf {\bibinfo {volume} {15}},\ \bibinfo {pages} {4961} (\bibinfo {year} {2024})}\BibitemShut {NoStop}%
\bibitem [{\citenamefont {Feng}\ \emph {et~al.}(2022)\citenamefont {Feng}, \citenamefont {Zhou}, \citenamefont {{Šmejkal}}, \citenamefont {Wu}, \citenamefont {Zhu}, \citenamefont {Guo}, \citenamefont {Gonz\'alez-Hern\'andez}, \citenamefont {Wang}, \citenamefont {Yan}, \citenamefont {Qin}, \citenamefont {Zhang}, \citenamefont {Wu}, \citenamefont {Chen}, \citenamefont {Meng}, \citenamefont {Liu}, \citenamefont {Xia}, \citenamefont {Sinova}, \citenamefont {Jungwirth},\ and\ \citenamefont {Liu}}]{Feng_NatElectron2022}%
  \BibitemOpen
  \bibfield  {author} {\bibinfo {author} {\bibfnamefont {Z.}~\bibnamefont {Feng}}, \bibinfo {author} {\bibfnamefont {X.}~\bibnamefont {Zhou}}, \bibinfo {author} {\bibfnamefont {L.}~\bibnamefont {{Šmejkal}}}, \bibinfo {author} {\bibfnamefont {L.}~\bibnamefont {Wu}}, \bibinfo {author} {\bibfnamefont {Z.}~\bibnamefont {Zhu}}, \bibinfo {author} {\bibfnamefont {H.}~\bibnamefont {Guo}}, \bibinfo {author} {\bibfnamefont {R.}~\bibnamefont {Gonz\'alez-Hern\'andez}}, \bibinfo {author} {\bibfnamefont {X.}~\bibnamefont {Wang}}, \bibinfo {author} {\bibfnamefont {H.}~\bibnamefont {Yan}}, \bibinfo {author} {\bibfnamefont {P.}~\bibnamefont {Qin}}, \bibinfo {author} {\bibfnamefont {X.}~\bibnamefont {Zhang}}, \bibinfo {author} {\bibfnamefont {H.}~\bibnamefont {Wu}}, \bibinfo {author} {\bibfnamefont {H.}~\bibnamefont {Chen}}, \bibinfo {author} {\bibfnamefont {Z.}~\bibnamefont {Meng}}, \bibinfo {author} {\bibfnamefont {L.}~\bibnamefont {Liu}}, \bibinfo {author} {\bibfnamefont {Z.}~\bibnamefont {Xia}}, \bibinfo {author}
  {\bibfnamefont {J.}~\bibnamefont {Sinova}}, \bibinfo {author} {\bibfnamefont {T.}~\bibnamefont {Jungwirth}}, \ and\ \bibinfo {author} {\bibfnamefont {Z.}~\bibnamefont {Liu}},\ }\bibfield  {title} {\enquote {\bibinfo {title} {An anomalous {H}all effect in altermagnetic ruthenium dioxide},}\ }\href {\doibase 10.1038/s41928-022-00866-z} {\bibfield  {journal} {\bibinfo  {journal} {Nat. Electron.}\ }\textbf {\bibinfo {volume} {5}},\ \bibinfo {pages} {735--743} (\bibinfo {year} {2022})}\BibitemShut {NoStop}%
\bibitem [{\citenamefont {Sheoran}\ and\ \citenamefont {Dev}(2025)}]{Sheoran_PRB2025}%
  \BibitemOpen
  \bibfield  {author} {\bibinfo {author} {\bibfnamefont {S.}~\bibnamefont {Sheoran}}\ and\ \bibinfo {author} {\bibfnamefont {P.}~\bibnamefont {Dev}},\ }\bibfield  {title} {\enquote {\bibinfo {title} {Spontaneous anomalous {H}all effect in two-dimensional altermagnets},}\ }\href {\doibase 10.1103/PhysRevB.111.184407} {\bibfield  {journal} {\bibinfo  {journal} {Phys. Rev. B}\ }\textbf {\bibinfo {volume} {111}},\ \bibinfo {pages} {184407} (\bibinfo {year} {2025})}\BibitemShut {NoStop}%
\bibitem [{\citenamefont {Yu}\ \emph {et~al.}(2025)\citenamefont {Yu}, \citenamefont {Shahid},\ and\ \citenamefont {Liu}}]{Yu_npjQuantumMater2025}%
  \BibitemOpen
  \bibfield  {author} {\bibinfo {author} {\bibfnamefont {T.}~\bibnamefont {Yu}}, \bibinfo {author} {\bibfnamefont {I.}~\bibnamefont {Shahid}}, \ and\ \bibinfo {author} {\bibfnamefont {P.}~\bibnamefont {Liu}},\ }\bibfield  {title} {\enquote {\bibinfo {title} {N\'eel vector-dependent anomalous transport in altermagnetic metal {C}r{S}b},}\ }\href {\doibase 10.1038/s41535-025-00766-3} {\bibfield  {journal} {\bibinfo  {journal} {npj Quantum Mater.}\ }\textbf {\bibinfo {volume} {10}},\ \bibinfo {pages} {47} (\bibinfo {year} {2025})}\BibitemShut {NoStop}%
\bibitem [{\citenamefont {{Šmejkal}}\ \emph {et~al.}(2020)\citenamefont {{Šmejkal}}, \citenamefont {Gonz\'alez-Hern\'andez}, \citenamefont {Jungwirth},\ and\ \citenamefont {Sinova}}]{Smejkal_SciAdv2020}%
  \BibitemOpen
  \bibfield  {author} {\bibinfo {author} {\bibfnamefont {L.}~\bibnamefont {{Šmejkal}}}, \bibinfo {author} {\bibfnamefont {R.}~\bibnamefont {Gonz\'alez-Hern\'andez}}, \bibinfo {author} {\bibfnamefont {T.}~\bibnamefont {Jungwirth}}, \ and\ \bibinfo {author} {\bibfnamefont {J.}~\bibnamefont {Sinova}},\ }\bibfield  {title} {\enquote {\bibinfo {title} {Crystal time-reversal symmetry breaking and spontaneous {H}all effect in collinear antiferromagnets},}\ }\href {\doibase 10.1126/sciadv.aaz8809} {\bibfield  {journal} {\bibinfo  {journal} {Sci. Adv.}\ }\textbf {\bibinfo {volume} {6}},\ \bibinfo {pages} {eaaz8809} (\bibinfo {year} {2020})}\BibitemShut {NoStop}%
\bibitem [{\citenamefont {Farajollahpour}\ \emph {et~al.}(2025)\citenamefont {Farajollahpour}, \citenamefont {Ganesh},\ and\ \citenamefont {Samokhin}}]{Farajollahpour_npj2025}%
  \BibitemOpen
  \bibfield  {author} {\bibinfo {author} {\bibfnamefont {T.}~\bibnamefont {Farajollahpour}}, \bibinfo {author} {\bibfnamefont {R.}~\bibnamefont {Ganesh}}, \ and\ \bibinfo {author} {\bibfnamefont {K.~V.}\ \bibnamefont {Samokhin}},\ }\bibfield  {title} {\enquote {\bibinfo {title} {{B}erry curvature-induced transport signature for altermagnetic order},}\ }\href {\doibase 10.1038/s41535-025-00805-zw} {\bibfield  {journal} {\bibinfo  {journal} {npj Quantum Mater.}\ }\textbf {\bibinfo {volume} {10}},\ \bibinfo {pages} {77} (\bibinfo {year} {2025})}\BibitemShut {NoStop}%
\bibitem [{\citenamefont {Bo}\ \emph {et~al.}(2024)\citenamefont {Bo}, \citenamefont {Zhou},\ and\ \citenamefont {Sun}}]{Bo_PRB2024}%
  \BibitemOpen
  \bibfield  {author} {\bibinfo {author} {\bibfnamefont {X.}~\bibnamefont {Bo}}, \bibinfo {author} {\bibfnamefont {J.}~\bibnamefont {Zhou}}, \ and\ \bibinfo {author} {\bibfnamefont {Q.}~\bibnamefont {Sun}},\ }\bibfield  {title} {\enquote {\bibinfo {title} {Crystal {H}all effect and {B}erry curvature dipole in altermagnets},}\ }\href {\doibase 10.1103/PhysRevB.109.035117} {\bibfield  {journal} {\bibinfo  {journal} {Phys. Rev. B}\ }\textbf {\bibinfo {volume} {109}},\ \bibinfo {pages} {035117} (\bibinfo {year} {2024})}\BibitemShut {NoStop}%
\bibitem [{\citenamefont {Hariki}\ and\ \citenamefont {Kune\v{s}}(2023)}]{Hariki_PRB2023}%
  \BibitemOpen
  \bibfield  {author} {\bibinfo {author} {\bibfnamefont {A.}~\bibnamefont {Hariki}}\ and\ \bibinfo {author} {\bibfnamefont {J.}~\bibnamefont {Kune\v{s}}},\ }\bibfield  {title} {\enquote {\bibinfo {title} {Anisotropic optical conductivity in altermagnetic semimetals},}\ }\href {\doibase 10.1103/PhysRevB.107.125123} {\bibfield  {journal} {\bibinfo  {journal} {Phys. Rev. B}\ }\textbf {\bibinfo {volume} {107}},\ \bibinfo {pages} {125123} (\bibinfo {year} {2023})}\BibitemShut {NoStop}%
\bibitem [{\citenamefont {Gonz\'alez-Hern\'andez}\ \emph {et~al.}(2025)\citenamefont {Gonz\'alez-Hern\'andez}, \citenamefont {Serrano},\ and\ \citenamefont {Uribe}}]{Gonzalez_PRB2025}%
  \BibitemOpen
  \bibfield  {author} {\bibinfo {author} {\bibfnamefont {R.}~\bibnamefont {Gonz\'alez-Hern\'andez}}, \bibinfo {author} {\bibfnamefont {H.}~\bibnamefont {Serrano}}, \ and\ \bibinfo {author} {\bibfnamefont {B.}~\bibnamefont {Uribe}},\ }\bibfield  {title} {\enquote {\bibinfo {title} {Spin {C}hern number in altermagnets},}\ }\href {\doibase 10.1103/PhysRevB.111.085127} {\bibfield  {journal} {\bibinfo  {journal} {Phys. Rev. B}\ }\textbf {\bibinfo {volume} {111}},\ \bibinfo {pages} {085127} (\bibinfo {year} {2025})}\BibitemShut {NoStop}%
\bibitem [{\citenamefont {Li}\ \emph {et~al.}(2024)\citenamefont {Li}, \citenamefont {Liu},\ and\ \citenamefont {Liu}}]{Yu-Xuan_PRB2024}%
  \BibitemOpen
  \bibfield  {author} {\bibinfo {author} {\bibfnamefont {Y.-X.}\ \bibnamefont {Li}}, \bibinfo {author} {\bibfnamefont {Y.}~\bibnamefont {Liu}}, \ and\ \bibinfo {author} {\bibfnamefont {C.-C.}\ \bibnamefont {Liu}},\ }\bibfield  {title} {\enquote {\bibinfo {title} {Creation and manipulation of higher-order topological states by altermagnets},}\ }\href {\doibase 10.1103/PhysRevB.109.L201109} {\bibfield  {journal} {\bibinfo  {journal} {Phys. Rev. B}\ }\textbf {\bibinfo {volume} {109}},\ \bibinfo {pages} {L201109} (\bibinfo {year} {2024})}\BibitemShut {NoStop}%
\bibitem [{\citenamefont {Brekke}\ \emph {et~al.}(2023)\citenamefont {Brekke}, \citenamefont {Brataas},\ and\ \citenamefont {Sudb{\o}}}]{Brekke_PRB2023}%
  \BibitemOpen
  \bibfield  {author} {\bibinfo {author} {\bibfnamefont {B.}~\bibnamefont {Brekke}}, \bibinfo {author} {\bibfnamefont {A.}~\bibnamefont {Brataas}}, \ and\ \bibinfo {author} {\bibfnamefont {A.}~\bibnamefont {Sudb{\o}}},\ }\bibfield  {title} {\enquote {\bibinfo {title} {Two-dimensional altermagnets: {S}uperconductivity in a minimal microscopic model},}\ }\href {\doibase 10.1103/PhysRevB.108.224421} {\bibfield  {journal} {\bibinfo  {journal} {Phys. Rev. B}\ }\textbf {\bibinfo {volume} {108}},\ \bibinfo {pages} {224421} (\bibinfo {year} {2023})}\BibitemShut {NoStop}%
\bibitem [{\citenamefont {Chen}\ \emph {et~al.}(2025)\citenamefont {Chen}, \citenamefont {Wang}, \citenamefont {Wu}, \citenamefont {Sun}, \citenamefont {Zhou}, \citenamefont {Wang},\ and\ \citenamefont {Xu}}]{RuiChen_PRL2025}%
  \BibitemOpen
  \bibfield  {author} {\bibinfo {author} {\bibfnamefont {R.}~\bibnamefont {Chen}}, \bibinfo {author} {\bibfnamefont {Z.-M.}\ \bibnamefont {Wang}}, \bibinfo {author} {\bibfnamefont {K.}~\bibnamefont {Wu}}, \bibinfo {author} {\bibfnamefont {H.-P.}\ \bibnamefont {Sun}}, \bibinfo {author} {\bibfnamefont {B.}~\bibnamefont {Zhou}}, \bibinfo {author} {\bibfnamefont {R.}~\bibnamefont {Wang}}, \ and\ \bibinfo {author} {\bibfnamefont {D.-H.}\ \bibnamefont {Xu}},\ }\bibfield  {title} {\enquote {\bibinfo {title} {Probing $\mathbit{k}$-space alternating spin polarization via the anomalous {H}all effect},}\ }\href {\doibase 10.1103/yrs7-m6zy} {\bibfield  {journal} {\bibinfo  {journal} {Phys. Rev. Lett.}\ }\textbf {\bibinfo {volume} {135}},\ \bibinfo {pages} {096602} (\bibinfo {year} {2025})}\BibitemShut {NoStop}%
\bibitem [{\citenamefont {Hasan}\ and\ \citenamefont {Kane}(2010)}]{Hasan_RMP2010}%
  \BibitemOpen
  \bibfield  {author} {\bibinfo {author} {\bibfnamefont {M.~Z.}\ \bibnamefont {Hasan}}\ and\ \bibinfo {author} {\bibfnamefont {C.~L.}\ \bibnamefont {Kane}},\ }\bibfield  {title} {\enquote {\bibinfo {title} {Colloquium: Topological insulators},}\ }\href {\doibase 10.1103/RevModPhys.82.3045} {\bibfield  {journal} {\bibinfo  {journal} {Rev. Mod. Phys.}\ }\textbf {\bibinfo {volume} {82}},\ \bibinfo {pages} {3045--3067} (\bibinfo {year} {2010})}\BibitemShut {NoStop}%
\bibitem [{\citenamefont {Bernevig}\ \emph {et~al.}(2006)\citenamefont {Bernevig}, \citenamefont {Hughes},\ and\ \citenamefont {Zhang}}]{Bernevig_Science2006}%
  \BibitemOpen
  \bibfield  {author} {\bibinfo {author} {\bibfnamefont {B.~A.}\ \bibnamefont {Bernevig}}, \bibinfo {author} {\bibfnamefont {T.~L.}\ \bibnamefont {Hughes}}, \ and\ \bibinfo {author} {\bibfnamefont {S.-C.}\ \bibnamefont {Zhang}},\ }\bibfield  {title} {\enquote {\bibinfo {title} {Quantum spin {H}all effect and topological phase transition in {H}g{T}e quantum wells},}\ }\href {\doibase 10.1126/science.1133734} {\bibfield  {journal} {\bibinfo  {journal} {Science}\ }\textbf {\bibinfo {volume} {314}},\ \bibinfo {pages} {1757--1761} (\bibinfo {year} {2006})}\BibitemShut {NoStop}%
\bibitem [{\citenamefont {Qi}\ and\ \citenamefont {Zhang}(2011)}]{Qi_Zhang_RevModPhys2011}%
  \BibitemOpen
  \bibfield  {author} {\bibinfo {author} {\bibfnamefont {X.-L.}\ \bibnamefont {Qi}}\ and\ \bibinfo {author} {\bibfnamefont {S.-C.}\ \bibnamefont {Zhang}},\ }\bibfield  {title} {\enquote {\bibinfo {title} {Topological insulators and superconductors},}\ }\href {\doibase 10.1103/RevModPhys.83.1057} {\bibfield  {journal} {\bibinfo  {journal} {Rev. Mod. Phys.}\ }\textbf {\bibinfo {volume} {83}},\ \bibinfo {pages} {1057--1110} (\bibinfo {year} {2011})}\BibitemShut {NoStop}%
\bibitem [{\citenamefont {Hsieh}\ \emph {et~al.}(2009)\citenamefont {Hsieh}, \citenamefont {Xia}, \citenamefont {Qian}, \citenamefont {Wray}, \citenamefont {Dil}, \citenamefont {Meier}, \citenamefont {Osterwalder}, \citenamefont {Patthey}, \citenamefont {Fedorov}, \citenamefont {Lin}, \citenamefont {Bansil}, \citenamefont {Grauer}, \citenamefont {Xu},\ and\ \citenamefont {Hasan}}]{Hsieh_Nature2009}%
  \BibitemOpen
  \bibfield  {author} {\bibinfo {author} {\bibfnamefont {D.}~\bibnamefont {Hsieh}}, \bibinfo {author} {\bibfnamefont {Y.}~\bibnamefont {Xia}}, \bibinfo {author} {\bibfnamefont {D.}~\bibnamefont {Qian}}, \bibinfo {author} {\bibfnamefont {L.}~\bibnamefont {Wray}}, \bibinfo {author} {\bibfnamefont {J.~H.}\ \bibnamefont {Dil}}, \bibinfo {author} {\bibfnamefont {F.}~\bibnamefont {Meier}}, \bibinfo {author} {\bibfnamefont {J.}~\bibnamefont {Osterwalder}}, \bibinfo {author} {\bibfnamefont {L.}~\bibnamefont {Patthey}}, \bibinfo {author} {\bibfnamefont {A.~V.}\ \bibnamefont {Fedorov}}, \bibinfo {author} {\bibfnamefont {H.}~\bibnamefont {Lin}}, \bibinfo {author} {\bibfnamefont {A.}~\bibnamefont {Bansil}}, \bibinfo {author} {\bibfnamefont {D.}~\bibnamefont {Grauer}}, \bibinfo {author} {\bibfnamefont {S.-Y.}\ \bibnamefont {Xu}}, \ and\ \bibinfo {author} {\bibfnamefont {M.~Z.}\ \bibnamefont {Hasan}},\ }\bibfield  {title} {\enquote {\bibinfo {title} {A tunable topological insulator in the spin helical {D}irac transport
  regime},}\ }\href {\doibase 10.1038/nature08234} {\bibfield  {journal} {\bibinfo  {journal} {Nature}\ }\textbf {\bibinfo {volume} {460}},\ \bibinfo {pages} {1101--1105} (\bibinfo {year} {2009})}\BibitemShut {NoStop}%
\bibitem [{\citenamefont {Shi}\ \emph {et~al.}(2025)\citenamefont {Shi}, \citenamefont {Liu}, \citenamefont {Hu}, \citenamefont {Shi}, \citenamefont {Manchon},\ and\ \citenamefont {Yang}}]{Shi_PRB2025}%
  \BibitemOpen
  \bibfield  {author} {\bibinfo {author} {\bibfnamefont {S.}~\bibnamefont {Shi}}, \bibinfo {author} {\bibfnamefont {E.}~\bibnamefont {Liu}}, \bibinfo {author} {\bibfnamefont {F.}~\bibnamefont {Hu}}, \bibinfo {author} {\bibfnamefont {G.}~\bibnamefont {Shi}}, \bibinfo {author} {\bibfnamefont {A.}~\bibnamefont {Manchon}}, \ and\ \bibinfo {author} {\bibfnamefont {H.}~\bibnamefont {Yang}},\ }\bibfield  {title} {\enquote {\bibinfo {title} {Nonreciprocal spin-charge interconversion in topological insulator/ferromagnet heterostructures},}\ }\href {\doibase 10.1103/PhysRevB.111.094433} {\bibfield  {journal} {\bibinfo  {journal} {Phys. Rev. B}\ }\textbf {\bibinfo {volume} {111}},\ \bibinfo {pages} {094433} (\bibinfo {year} {2025})}\BibitemShut {NoStop}%
\bibitem [{\citenamefont {Nagaosa}\ \emph {et~al.}(2010)\citenamefont {Nagaosa}, \citenamefont {Sinova}, \citenamefont {Onoda}, \citenamefont {MacDonald},\ and\ \citenamefont {Ong}}]{Nagaosa_AHE_RMP2010}%
  \BibitemOpen
  \bibfield  {author} {\bibinfo {author} {\bibfnamefont {N.}~\bibnamefont {Nagaosa}}, \bibinfo {author} {\bibfnamefont {J.}~\bibnamefont {Sinova}}, \bibinfo {author} {\bibfnamefont {S.}~\bibnamefont {Onoda}}, \bibinfo {author} {\bibfnamefont {A.~H.}\ \bibnamefont {MacDonald}}, \ and\ \bibinfo {author} {\bibfnamefont {N.~P.}\ \bibnamefont {Ong}},\ }\bibfield  {title} {\enquote {\bibinfo {title} {Anomalous {H}all effect},}\ }\href {\doibase 10.1103/RevModPhys.82.1539} {\bibfield  {journal} {\bibinfo  {journal} {Rev. Mod. Phys.}\ }\textbf {\bibinfo {volume} {82}},\ \bibinfo {pages} {1539--1592} (\bibinfo {year} {2010})}\BibitemShut {NoStop}%
\bibitem [{\citenamefont {Chang}\ \emph {et~al.}(2023)\citenamefont {Chang}, \citenamefont {Liu},\ and\ \citenamefont {MacDonald}}]{Chang_RevModPhy2023}%
  \BibitemOpen
  \bibfield  {author} {\bibinfo {author} {\bibfnamefont {C.-Z.}\ \bibnamefont {Chang}}, \bibinfo {author} {\bibfnamefont {C.-X.}\ \bibnamefont {Liu}}, \ and\ \bibinfo {author} {\bibfnamefont {A.~H.}\ \bibnamefont {MacDonald}},\ }\bibfield  {title} {\enquote {\bibinfo {title} {Colloquium: Quantum anomalous {H}all effect},}\ }\href {\doibase 10.1103/RevModPhys.95.011002} {\bibfield  {journal} {\bibinfo  {journal} {Rev. Mod. Phys.}\ }\textbf {\bibinfo {volume} {95}},\ \bibinfo {pages} {011002} (\bibinfo {year} {2023})}\BibitemShut {NoStop}%
\bibitem [{\citenamefont {Bychkov}\ and\ \citenamefont {Rashba}(1984)}]{Rashba_1984}%
  \BibitemOpen
  \bibfield  {author} {\bibinfo {author} {\bibfnamefont {Y.~A.}\ \bibnamefont {Bychkov}}\ and\ \bibinfo {author} {\bibfnamefont {E.~I.}\ \bibnamefont {Rashba}},\ }\bibfield  {title} {\enquote {\bibinfo {title} {Oscillatory effects and the magnetic susceptibility of carriers in inversion layers},}\ }\href {\doibase 10.1088/0022-3719/17/33/015} {\bibfield  {journal} {\bibinfo  {journal} {J. Phys. C}\ }\textbf {\bibinfo {volume} {17}},\ \bibinfo {pages} {6039} (\bibinfo {year} {1984})}\BibitemShut {NoStop}%
\bibitem [{\citenamefont {Edelstein}(1990)}]{Edelstein_SSC1990}%
  \BibitemOpen
  \bibfield  {author} {\bibinfo {author} {\bibfnamefont {V.~M.}\ \bibnamefont {Edelstein}},\ }\bibfield  {title} {\enquote {\bibinfo {title} {Spin polarization of conduction electrons induced by electric current in two-dimensional asymmetric electron systems},}\ }\href {\doibase 10.1016/0038-1098(90)90963-C} {\bibfield  {journal} {\bibinfo  {journal} {Solid State Commun.}\ }\textbf {\bibinfo {volume} {73}},\ \bibinfo {pages} {233--235} (\bibinfo {year} {1990})}\BibitemShut {NoStop}%
\bibitem [{\citenamefont {Rojas-S\'a{}nchez}\ \emph {et~al.}(2013)\citenamefont {Rojas-S\'a{}nchez}, \citenamefont {Vila}, \citenamefont {Desfonds}, \citenamefont {Gambarelli}, \citenamefont {Attan\'e}, \citenamefont {De~Teresa}, \citenamefont {Mag\'en},\ and\ \citenamefont {Fert}}]{Rojas_NatCommun2013}%
  \BibitemOpen
  \bibfield  {author} {\bibinfo {author} {\bibfnamefont {J.~C.}\ \bibnamefont {Rojas-S\'a{}nchez}}, \bibinfo {author} {\bibfnamefont {L.}~\bibnamefont {Vila}}, \bibinfo {author} {\bibfnamefont {G.}~\bibnamefont {Desfonds}}, \bibinfo {author} {\bibfnamefont {S.}~\bibnamefont {Gambarelli}}, \bibinfo {author} {\bibfnamefont {J.~P.}\ \bibnamefont {Attan\'e}}, \bibinfo {author} {\bibfnamefont {J.~M.}\ \bibnamefont {De~Teresa}}, \bibinfo {author} {\bibfnamefont {C.}~\bibnamefont {Mag\'en}}, \ and\ \bibinfo {author} {\bibfnamefont {A.}~\bibnamefont {Fert}},\ }\bibfield  {title} {\enquote {\bibinfo {title} {Spin-to-charge conversion using {R}ashba coupling at the interface between non-magnetic materials},}\ }\href {\doibase 10.1038/ncomms3944} {\bibfield  {journal} {\bibinfo  {journal} {Nat. Commun.}\ }\textbf {\bibinfo {volume} {4}},\ \bibinfo {pages} {2944} (\bibinfo {year} {2013})}\BibitemShut {NoStop}%
\bibitem [{\citenamefont {Lesne}\ \emph {et~al.}(2016)\citenamefont {Lesne}, \citenamefont {Fu}, \citenamefont {Oyarzun}, \citenamefont {Rojas-S\'a{}nchez}, \citenamefont {Vaz}, \citenamefont {Naganuma}, \citenamefont {Sicoli}, \citenamefont {Pizzini}, \citenamefont {Youssef}, \citenamefont {Barthel}, \citenamefont {Fert}, \citenamefont {Bibes}, \citenamefont {Barth\'el\'emy},\ and\ \citenamefont {Vila}}]{Lesne_NatMater2016}%
  \BibitemOpen
  \bibfield  {author} {\bibinfo {author} {\bibfnamefont {E.}~\bibnamefont {Lesne}}, \bibinfo {author} {\bibfnamefont {Y.}~\bibnamefont {Fu}}, \bibinfo {author} {\bibfnamefont {S.}~\bibnamefont {Oyarzun}}, \bibinfo {author} {\bibfnamefont {J.~C.}\ \bibnamefont {Rojas-S\'a{}nchez}}, \bibinfo {author} {\bibfnamefont {D.~C.}\ \bibnamefont {Vaz}}, \bibinfo {author} {\bibfnamefont {H.}~\bibnamefont {Naganuma}}, \bibinfo {author} {\bibfnamefont {G.}~\bibnamefont {Sicoli}}, \bibinfo {author} {\bibfnamefont {S.}~\bibnamefont {Pizzini}}, \bibinfo {author} {\bibfnamefont {J.~B.}\ \bibnamefont {Youssef}}, \bibinfo {author} {\bibfnamefont {J.}~\bibnamefont {Barthel}}, \bibinfo {author} {\bibfnamefont {A.}~\bibnamefont {Fert}}, \bibinfo {author} {\bibfnamefont {M.}~\bibnamefont {Bibes}}, \bibinfo {author} {\bibfnamefont {A.}~\bibnamefont {Barth\'el\'emy}}, \ and\ \bibinfo {author} {\bibfnamefont {L.}~\bibnamefont {Vila}},\ }\bibfield  {title} {\enquote {\bibinfo {title} {Highly efficient and tunable spin-to-charge
  conversion through {R}ashba coupling at oxide interfaces},}\ }\href {\doibase 10.1038/nmat4726} {\bibfield  {journal} {\bibinfo  {journal} {Nat. Mater.}\ }\textbf {\bibinfo {volume} {15}},\ \bibinfo {pages} {1261--1266} (\bibinfo {year} {2016})}\BibitemShut {NoStop}%
\bibitem [{\citenamefont {Mellnik}\ \emph {et~al.}(2014)\citenamefont {Mellnik}, \citenamefont {Lee}, \citenamefont {Richardella}, \citenamefont {Grab}, \citenamefont {Mintun}, \citenamefont {Fischer}, \citenamefont {Vaezi}, \citenamefont {Manchon}, \citenamefont {Kim}, \citenamefont {Samarth},\ and\ \citenamefont {Ralph}}]{Mellnik_Nature2014}%
  \BibitemOpen
  \bibfield  {author} {\bibinfo {author} {\bibfnamefont {J.~R.}\ \bibnamefont {Mellnik}}, \bibinfo {author} {\bibfnamefont {J.~S.}\ \bibnamefont {Lee}}, \bibinfo {author} {\bibfnamefont {A.}~\bibnamefont {Richardella}}, \bibinfo {author} {\bibfnamefont {J.~L.}\ \bibnamefont {Grab}}, \bibinfo {author} {\bibfnamefont {P.~J.}\ \bibnamefont {Mintun}}, \bibinfo {author} {\bibfnamefont {M.~H.}\ \bibnamefont {Fischer}}, \bibinfo {author} {\bibfnamefont {A.}~\bibnamefont {Vaezi}}, \bibinfo {author} {\bibfnamefont {A.}~\bibnamefont {Manchon}}, \bibinfo {author} {\bibfnamefont {E.~A.}\ \bibnamefont {Kim}}, \bibinfo {author} {\bibfnamefont {N.}~\bibnamefont {Samarth}}, \ and\ \bibinfo {author} {\bibfnamefont {D.~C.}\ \bibnamefont {Ralph}},\ }\bibfield  {title} {\enquote {\bibinfo {title} {Spin-transfer torque generated by a topological insulator},}\ }\href {\doibase 10.1038/nature13534} {\bibfield  {journal} {\bibinfo  {journal} {Nature}\ }\textbf {\bibinfo {volume} {511}},\ \bibinfo {pages} {449--451} (\bibinfo {year}
  {2014})}\BibitemShut {NoStop}%
\bibitem [{\citenamefont {Liu}\ \emph {et~al.}(2018)\citenamefont {Liu}, \citenamefont {Besbas}, \citenamefont {Wang}, \citenamefont {He}, \citenamefont {Chen}, \citenamefont {Zhu}, \citenamefont {Wu}, \citenamefont {Lee}, \citenamefont {Wang}, \citenamefont {Moon}, \citenamefont {Koirala}, \citenamefont {Oh},\ and\ \citenamefont {Yang}}]{Liu_NatCommu2018}%
  \BibitemOpen
  \bibfield  {author} {\bibinfo {author} {\bibfnamefont {Y.}~\bibnamefont {Liu}}, \bibinfo {author} {\bibfnamefont {J.}~\bibnamefont {Besbas}}, \bibinfo {author} {\bibfnamefont {Y.}~\bibnamefont {Wang}}, \bibinfo {author} {\bibfnamefont {P.}~\bibnamefont {He}}, \bibinfo {author} {\bibfnamefont {M.}~\bibnamefont {Chen}}, \bibinfo {author} {\bibfnamefont {D.}~\bibnamefont {Zhu}}, \bibinfo {author} {\bibfnamefont {Y.}~\bibnamefont {Wu}}, \bibinfo {author} {\bibfnamefont {J.~M.}\ \bibnamefont {Lee}}, \bibinfo {author} {\bibfnamefont {L.}~\bibnamefont {Wang}}, \bibinfo {author} {\bibfnamefont {J.}~\bibnamefont {Moon}}, \bibinfo {author} {\bibfnamefont {N.}~\bibnamefont {Koirala}}, \bibinfo {author} {\bibfnamefont {S.}~\bibnamefont {Oh}}, \ and\ \bibinfo {author} {\bibfnamefont {H.}~\bibnamefont {Yang}},\ }\bibfield  {title} {\enquote {\bibinfo {title} {Direct visualization of current-induced spin accumulation in topological insulators},}\ }\href {\doibase 10.1038/s41467-018-04939-6} {\bibfield  {journal} {\bibinfo
   {journal} {Nat. Commun.}\ }\textbf {\bibinfo {volume} {9}},\ \bibinfo {pages} {2492} (\bibinfo {year} {2018})}\BibitemShut {NoStop}%
\bibitem [{\citenamefont {Shiomi}\ \emph {et~al.}(2014)\citenamefont {Shiomi}, \citenamefont {Nomura}, \citenamefont {Kajiwara}, \citenamefont {Eto}, \citenamefont {Novak}, \citenamefont {Segawa}, \citenamefont {Ando},\ and\ \citenamefont {Saitoh}}]{Shiomi_PRL2014}%
  \BibitemOpen
  \bibfield  {author} {\bibinfo {author} {\bibfnamefont {Y.}~\bibnamefont {Shiomi}}, \bibinfo {author} {\bibfnamefont {K.}~\bibnamefont {Nomura}}, \bibinfo {author} {\bibfnamefont {Y.}~\bibnamefont {Kajiwara}}, \bibinfo {author} {\bibfnamefont {K.}~\bibnamefont {Eto}}, \bibinfo {author} {\bibfnamefont {M.}~\bibnamefont {Novak}}, \bibinfo {author} {\bibfnamefont {K.}~\bibnamefont {Segawa}}, \bibinfo {author} {\bibfnamefont {Y.}~\bibnamefont {Ando}}, \ and\ \bibinfo {author} {\bibfnamefont {E.}~\bibnamefont {Saitoh}},\ }\bibfield  {title} {\enquote {\bibinfo {title} {Spin-electricity conversion induced by spin injection into topological insulators},}\ }\href {\doibase 10.1103/PhysRevLett.113.196601} {\bibfield  {journal} {\bibinfo  {journal} {Phys. Rev. Lett.}\ }\textbf {\bibinfo {volume} {113}},\ \bibinfo {pages} {196601} (\bibinfo {year} {2014})}\BibitemShut {NoStop}%
\bibitem [{\citenamefont {Bernevig}\ \emph {et~al.}(2022)\citenamefont {Bernevig}, \citenamefont {Felser},\ and\ \citenamefont {Beidenkopf}}]{Felser_NatMater2022}%
  \BibitemOpen
  \bibfield  {author} {\bibinfo {author} {\bibfnamefont {B.~A.}\ \bibnamefont {Bernevig}}, \bibinfo {author} {\bibfnamefont {C.}~\bibnamefont {Felser}}, \ and\ \bibinfo {author} {\bibfnamefont {H.}~\bibnamefont {Beidenkopf}},\ }\bibfield  {title} {\enquote {\bibinfo {title} {Progress and prospects in magnetic topological materials},}\ }\href {\doibase 10.1038/s41586-021-04105-x} {\bibfield  {journal} {\bibinfo  {journal} {Nature}\ }\textbf {\bibinfo {volume} {603}},\ \bibinfo {pages} {41--51} (\bibinfo {year} {2022})}\BibitemShut {NoStop}%
\bibitem [{\citenamefont {MacNeill}\ \emph {et~al.}(2017)\citenamefont {MacNeill}, \citenamefont {Stiehl}, \citenamefont {Guimar\~{a}es}, \citenamefont {Buhrman}, \citenamefont {Park},\ and\ \citenamefont {Ralph}}]{MacNeill_PRL2017}%
  \BibitemOpen
  \bibfield  {author} {\bibinfo {author} {\bibfnamefont {D.}~\bibnamefont {MacNeill}}, \bibinfo {author} {\bibfnamefont {G.~M.}\ \bibnamefont {Stiehl}}, \bibinfo {author} {\bibfnamefont {M.~H.~D.}\ \bibnamefont {Guimar\~{a}es}}, \bibinfo {author} {\bibfnamefont {R.~A.}\ \bibnamefont {Buhrman}}, \bibinfo {author} {\bibfnamefont {J.}~\bibnamefont {Park}}, \ and\ \bibinfo {author} {\bibfnamefont {D.~C.}\ \bibnamefont {Ralph}},\ }\bibfield  {title} {\enquote {\bibinfo {title} {Control of spin--orbit torques through crystal symmetry in {WT}e$_2$/ferromagnet bilayers},}\ }\href {\doibase 10.1038/nphys3933} {\bibfield  {journal} {\bibinfo  {journal} {Nat. Phys.}\ }\textbf {\bibinfo {volume} {13}},\ \bibinfo {pages} {300--305} (\bibinfo {year} {2017})}\BibitemShut {NoStop}%
\bibitem [{\citenamefont {Manchon}\ \emph {et~al.}(2019)\citenamefont {Manchon}, \citenamefont {{\v Z}elezn\'y}, \citenamefont {Miron}, \citenamefont {Jungwirth}, \citenamefont {Sinova}, \citenamefont {Thiaville}, \citenamefont {Garello},\ and\ \citenamefont {Gambardella}}]{Manchon_RMP2019}%
  \BibitemOpen
  \bibfield  {author} {\bibinfo {author} {\bibfnamefont {A.}~\bibnamefont {Manchon}}, \bibinfo {author} {\bibfnamefont {J.}~\bibnamefont {{\v Z}elezn\'y}}, \bibinfo {author} {\bibfnamefont {I.~M.}\ \bibnamefont {Miron}}, \bibinfo {author} {\bibfnamefont {T.}~\bibnamefont {Jungwirth}}, \bibinfo {author} {\bibfnamefont {J.}~\bibnamefont {Sinova}}, \bibinfo {author} {\bibfnamefont {A.}~\bibnamefont {Thiaville}}, \bibinfo {author} {\bibfnamefont {K.}~\bibnamefont {Garello}}, \ and\ \bibinfo {author} {\bibfnamefont {P.}~\bibnamefont {Gambardella}},\ }\bibfield  {title} {\enquote {\bibinfo {title} {Current-induced spin-orbit torques in ferromagnetic and antiferromagnetic systems},}\ }\href {\doibase 10.1103/RevModPhys.91.035004} {\bibfield  {journal} {\bibinfo  {journal} {Rev. Mod. Phys.}\ }\textbf {\bibinfo {volume} {91}},\ \bibinfo {pages} {035004} (\bibinfo {year} {2019})}\BibitemShut {NoStop}%
\bibitem [{\citenamefont {Seibold}\ \emph {et~al.}(2017)\citenamefont {Seibold}, \citenamefont {Caprara}, \citenamefont {Grilli},\ and\ \citenamefont {Raimondi}}]{Seibold_PRL2017}%
  \BibitemOpen
  \bibfield  {author} {\bibinfo {author} {\bibfnamefont {G.}~\bibnamefont {Seibold}}, \bibinfo {author} {\bibfnamefont {S.}~\bibnamefont {Caprara}}, \bibinfo {author} {\bibfnamefont {M.}~\bibnamefont {Grilli}}, \ and\ \bibinfo {author} {\bibfnamefont {R.}~\bibnamefont {Raimondi}},\ }\bibfield  {title} {\enquote {\bibinfo {title} {Theory of the spin galvanic effect at oxide interfaces},}\ }\href {\doibase 10.1103/PhysRevLett.119.256801} {\bibfield  {journal} {\bibinfo  {journal} {Phys. Rev. Lett.}\ }\textbf {\bibinfo {volume} {119}},\ \bibinfo {pages} {256801} (\bibinfo {year} {2017})}\BibitemShut {NoStop}%
\bibitem [{\citenamefont {Guillet}\ \emph {et~al.}(2020)\citenamefont {Guillet}, \citenamefont {Zucchetti}, \citenamefont {Marchionni}, \citenamefont {Hallal}, \citenamefont {Biagioni}, \citenamefont {Vergnaud}, \citenamefont {Marty}, \citenamefont {Okuno}, \citenamefont {Masseboeuf}, \citenamefont {Finazzi}, \citenamefont {Ciccacci}, \citenamefont {Chshiev}, \citenamefont {Bottegoni},\ and\ \citenamefont {Jamet}}]{Guillet_PRB2020}%
  \BibitemOpen
  \bibfield  {author} {\bibinfo {author} {\bibfnamefont {T.}~\bibnamefont {Guillet}}, \bibinfo {author} {\bibfnamefont {C.}~\bibnamefont {Zucchetti}}, \bibinfo {author} {\bibfnamefont {A.}~\bibnamefont {Marchionni}}, \bibinfo {author} {\bibfnamefont {A.}~\bibnamefont {Hallal}}, \bibinfo {author} {\bibfnamefont {P.}~\bibnamefont {Biagioni}}, \bibinfo {author} {\bibfnamefont {C.}~\bibnamefont {Vergnaud}}, \bibinfo {author} {\bibfnamefont {A.}~\bibnamefont {Marty}}, \bibinfo {author} {\bibfnamefont {H.}~\bibnamefont {Okuno}}, \bibinfo {author} {\bibfnamefont {A.}~\bibnamefont {Masseboeuf}}, \bibinfo {author} {\bibfnamefont {M.}~\bibnamefont {Finazzi}}, \bibinfo {author} {\bibfnamefont {F.}~\bibnamefont {Ciccacci}}, \bibinfo {author} {\bibfnamefont {M.}~\bibnamefont {Chshiev}}, \bibinfo {author} {\bibfnamefont {F.}~\bibnamefont {Bottegoni}}, \ and\ \bibinfo {author} {\bibfnamefont {M.}~\bibnamefont {Jamet}},\ }\bibfield  {title} {\enquote {\bibinfo {title} {Spin orbitronics at a topological
  insulator-semiconductor interface},}\ }\href {\doibase 10.1103/PhysRevB.101.184406} {\bibfield  {journal} {\bibinfo  {journal} {Phys. Rev. B}\ }\textbf {\bibinfo {volume} {101}},\ \bibinfo {pages} {184406} (\bibinfo {year} {2020})}\BibitemShut {NoStop}%
\bibitem [{\citenamefont {Kondou}\ \emph {et~al.}(2016)\citenamefont {Kondou}, \citenamefont {Yoshimi}, \citenamefont {Tsukazaki}, \citenamefont {Fukuma}, \citenamefont {Matsuno}, \citenamefont {Takahashi}, \citenamefont {Kawasaki}, \citenamefont {Tokura},\ and\ \citenamefont {Y.}}]{Kondou_NatPhys2016}%
  \BibitemOpen
  \bibfield  {author} {\bibinfo {author} {\bibfnamefont {K.}~\bibnamefont {Kondou}}, \bibinfo {author} {\bibfnamefont {R.}~\bibnamefont {Yoshimi}}, \bibinfo {author} {\bibfnamefont {A.}~\bibnamefont {Tsukazaki}}, \bibinfo {author} {\bibfnamefont {Y.}~\bibnamefont {Fukuma}}, \bibinfo {author} {\bibfnamefont {J.}~\bibnamefont {Matsuno}}, \bibinfo {author} {\bibfnamefont {K.~S.}\ \bibnamefont {Takahashi}}, \bibinfo {author} {\bibfnamefont {M.}~\bibnamefont {Kawasaki}}, \bibinfo {author} {\bibfnamefont {Y.}~\bibnamefont {Tokura}}, \ and\ \bibinfo {author} {\bibfnamefont {Otani}\ \bibnamefont {Y.}},\ }\bibfield  {title} {\enquote {\bibinfo {title} {Fermi-level-dependent charge-to-spin current conversion by {D}irac surface states of topological insulators},}\ }\href {\doibase 10.1038/nphys3833} {\bibfield  {journal} {\bibinfo  {journal} {Nat. Phys.}\ }\textbf {\bibinfo {volume} {12}},\ \bibinfo {pages} {1027--1031} (\bibinfo {year} {2016})}\BibitemShut {NoStop}%
\bibitem [{\citenamefont {Dey}\ \emph {et~al.}(2018)\citenamefont {Dey}, \citenamefont {Prasad}, \citenamefont {Register},\ and\ \citenamefont {Banerjee}}]{DeyRik_PRB2018}%
  \BibitemOpen
  \bibfield  {author} {\bibinfo {author} {\bibfnamefont {R.}~\bibnamefont {Dey}}, \bibinfo {author} {\bibfnamefont {N.}~\bibnamefont {Prasad}}, \bibinfo {author} {\bibfnamefont {L.~F.}\ \bibnamefont {Register}}, \ and\ \bibinfo {author} {\bibfnamefont {S.~K.}\ \bibnamefont {Banerjee}},\ }\bibfield  {title} {\enquote {\bibinfo {title} {Conversion of spin current into charge current in a topological insulator: Role of the interface},}\ }\href {\doibase 10.1103/PhysRevB.97.174406} {\bibfield  {journal} {\bibinfo  {journal} {Phys. Rev. B}\ }\textbf {\bibinfo {volume} {97}},\ \bibinfo {pages} {174406} (\bibinfo {year} {2018})}\BibitemShut {NoStop}%
\bibitem [{\citenamefont {He}\ \emph {et~al.}(2021)\citenamefont {He}, \citenamefont {Tai}, \citenamefont {Wu}, \citenamefont {Wu}, \citenamefont {Razavi}, \citenamefont {Gosavi}, \citenamefont {Walker}, \citenamefont {Oguz}, \citenamefont {Lin}, \citenamefont {Wong}, \citenamefont {Liu}, \citenamefont {Dai},\ and\ \citenamefont {Wang}}]{Haoran_PRB2021}%
  \BibitemOpen
  \bibfield  {author} {\bibinfo {author} {\bibfnamefont {H.}~\bibnamefont {He}}, \bibinfo {author} {\bibfnamefont {L.}~\bibnamefont {Tai}}, \bibinfo {author} {\bibfnamefont {H.}~\bibnamefont {Wu}}, \bibinfo {author} {\bibfnamefont {D.}~\bibnamefont {Wu}}, \bibinfo {author} {\bibfnamefont {A.}~\bibnamefont {Razavi}}, \bibinfo {author} {\bibfnamefont {T.~A.}\ \bibnamefont {Gosavi}}, \bibinfo {author} {\bibfnamefont {E.~S.}\ \bibnamefont {Walker}}, \bibinfo {author} {\bibfnamefont {K.}~\bibnamefont {Oguz}}, \bibinfo {author} {\bibfnamefont {C.-C.}\ \bibnamefont {Lin}}, \bibinfo {author} {\bibfnamefont {K.}~\bibnamefont {Wong}}, \bibinfo {author} {\bibfnamefont {Y.}~\bibnamefont {Liu}}, \bibinfo {author} {\bibfnamefont {B.}~\bibnamefont {Dai}}, \ and\ \bibinfo {author} {\bibfnamefont {K.~L.}\ \bibnamefont {Wang}},\ }\bibfield  {title} {\enquote {\bibinfo {title} {Conversion between spin and charge currents in topological-insulator/nonmagnetic-metal systems},}\ }\href {\doibase 10.1103/PhysRevB.104.L220407}
  {\bibfield  {journal} {\bibinfo  {journal} {Phys. Rev. B}\ }\textbf {\bibinfo {volume} {104}},\ \bibinfo {pages} {L220407} (\bibinfo {year} {2021})}\BibitemShut {NoStop}%
\bibitem [{\citenamefont {Leiva-Montecinos}\ \emph {et~al.}(2023)\citenamefont {Leiva-Montecinos}, \citenamefont {Henk}, \citenamefont {Mertig},\ and\ \citenamefont {Johansson}}]{Leiva_PRevRes2023}%
  \BibitemOpen
  \bibfield  {author} {\bibinfo {author} {\bibfnamefont {S.}~\bibnamefont {Leiva-Montecinos}}, \bibinfo {author} {\bibfnamefont {J.}~\bibnamefont {Henk}}, \bibinfo {author} {\bibfnamefont {I.}~\bibnamefont {Mertig}}, \ and\ \bibinfo {author} {\bibfnamefont {A.}~\bibnamefont {Johansson}},\ }\bibfield  {title} {\enquote {\bibinfo {title} {Spin and orbital {E}delstein effect in a bilayer system with {R}ashba interaction},}\ }\href {\doibase 10.1103/PhysRevResearch.5.043294} {\bibfield  {journal} {\bibinfo  {journal} {Phys. Rev. Res.}\ }\textbf {\bibinfo {volume} {5}},\ \bibinfo {pages} {043294} (\bibinfo {year} {2023})}\BibitemShut {NoStop}%
\bibitem [{\citenamefont {Johansson}\ \emph {et~al.}(2016)\citenamefont {Johansson}, \citenamefont {Henk},\ and\ \citenamefont {Mertig}}]{Annika_PRB2016}%
  \BibitemOpen
  \bibfield  {author} {\bibinfo {author} {\bibfnamefont {A.}~\bibnamefont {Johansson}}, \bibinfo {author} {\bibfnamefont {J.}~\bibnamefont {Henk}}, \ and\ \bibinfo {author} {\bibfnamefont {I.}~\bibnamefont {Mertig}},\ }\bibfield  {title} {\enquote {\bibinfo {title} {Theoretical aspects of the edelstein effect for anisotropic two-dimensional electron gas and topological insulators},}\ }\href {\doibase 10.1103/PhysRevB.93.195440} {\bibfield  {journal} {\bibinfo  {journal} {Phys. Rev. B}\ }\textbf {\bibinfo {volume} {93}},\ \bibinfo {pages} {195440} (\bibinfo {year} {2016})}\BibitemShut {NoStop}%
\bibitem [{\citenamefont {Fan}\ \emph {et~al.}(2014)\citenamefont {Fan}, \citenamefont {Upadhyaya}, \citenamefont {Kou}, \citenamefont {Lang}, \citenamefont {Takei}, \citenamefont {Wang}, \citenamefont {Tang}, \citenamefont {He}, \citenamefont {Chang}, \citenamefont {Montazeri}, \citenamefont {Yu}, \citenamefont {Jiang}, \citenamefont {Nie}, \citenamefont {Schwartz}, \citenamefont {Wang},\ and\ \citenamefont {Wang}}]{Fan_NatMater2014}%
  \BibitemOpen
  \bibfield  {author} {\bibinfo {author} {\bibfnamefont {Y.}~\bibnamefont {Fan}}, \bibinfo {author} {\bibfnamefont {P.}~\bibnamefont {Upadhyaya}}, \bibinfo {author} {\bibfnamefont {X.}~\bibnamefont {Kou}}, \bibinfo {author} {\bibfnamefont {M.}~\bibnamefont {Lang}}, \bibinfo {author} {\bibfnamefont {S.}~\bibnamefont {Takei}}, \bibinfo {author} {\bibfnamefont {Z.}~\bibnamefont {Wang}}, \bibinfo {author} {\bibfnamefont {J.}~\bibnamefont {Tang}}, \bibinfo {author} {\bibfnamefont {L.}~\bibnamefont {He}}, \bibinfo {author} {\bibfnamefont {L.}~\bibnamefont {Chang}}, \bibinfo {author} {\bibfnamefont {M.}~\bibnamefont {Montazeri}}, \bibinfo {author} {\bibfnamefont {G.}~\bibnamefont {Yu}}, \bibinfo {author} {\bibfnamefont {W.}~\bibnamefont {Jiang}}, \bibinfo {author} {\bibfnamefont {T.}~\bibnamefont {Nie}}, \bibinfo {author} {\bibfnamefont {R.~N.}\ \bibnamefont {Schwartz}}, \bibinfo {author} {\bibfnamefont {Y.~T.}\ \bibnamefont {Wang}}, \ and\ \bibinfo {author} {\bibfnamefont {K.~L.}\ \bibnamefont {Wang}},\ }\bibfield
  {title} {\enquote {\bibinfo {title} {Magnetization switching through giant spin--orbit torque in a magnetically doped topological insulator},}\ }\href {\doibase 10.1038/nmat3973} {\bibfield  {journal} {\bibinfo  {journal} {Nat. Mater.}\ }\textbf {\bibinfo {volume} {13}},\ \bibinfo {pages} {699--704} (\bibinfo {year} {2014})}\BibitemShut {NoStop}%
\bibitem [{\citenamefont {Zelezn\'y}\ \emph {et~al.}(2014)\citenamefont {Zelezn\'y}, \citenamefont {Gao}, \citenamefont {V\'yborn\'y}, \citenamefont {Zemen}, \citenamefont {Ma{\v{s}}ek}, \citenamefont {Manchon}, \citenamefont {Wunderlich}, \citenamefont {Sinova},\ and\ \citenamefont {Jungwirth}}]{Zelezny_PRL2014}%
  \BibitemOpen
  \bibfield  {author} {\bibinfo {author} {\bibfnamefont {J.}~\bibnamefont {Zelezn\'y}}, \bibinfo {author} {\bibfnamefont {H.}~\bibnamefont {Gao}}, \bibinfo {author} {\bibfnamefont {K.}~\bibnamefont {V\'yborn\'y}}, \bibinfo {author} {\bibfnamefont {J.}~\bibnamefont {Zemen}}, \bibinfo {author} {\bibfnamefont {J.}~\bibnamefont {Ma{\v{s}}ek}}, \bibinfo {author} {\bibfnamefont {A.}~\bibnamefont {Manchon}}, \bibinfo {author} {\bibfnamefont {J.}~\bibnamefont {Wunderlich}}, \bibinfo {author} {\bibfnamefont {J.}~\bibnamefont {Sinova}}, \ and\ \bibinfo {author} {\bibfnamefont {T.}~\bibnamefont {Jungwirth}},\ }\bibfield  {title} {\enquote {\bibinfo {title} {Relativistic n{\'e}el-order fields induced by electrical current in antiferromagnets},}\ }\href {\doibase 10.1103/PhysRevLett.113.157201} {\bibfield  {journal} {\bibinfo  {journal} {Phys. Rev. Lett.}\ }\textbf {\bibinfo {volume} {113}},\ \bibinfo {pages} {157201} (\bibinfo {year} {2014})}\BibitemShut {NoStop}%
\bibitem [{\citenamefont {Prakash}\ \emph {et~al.}(2016)\citenamefont {Prakash}, \citenamefont {Brangham}, \citenamefont {Yang},\ and\ \citenamefont {Heremans}}]{Arati_PRB2016}%
  \BibitemOpen
  \bibfield  {author} {\bibinfo {author} {\bibfnamefont {A.}~\bibnamefont {Prakash}}, \bibinfo {author} {\bibfnamefont {J.}~\bibnamefont {Brangham}}, \bibinfo {author} {\bibfnamefont {F.}~\bibnamefont {Yang}}, \ and\ \bibinfo {author} {\bibfnamefont {J.~P.}\ \bibnamefont {Heremans}},\ }\bibfield  {title} {\enquote {\bibinfo {title} {Spin seebeck effect through antiferromagnetic {N}i{O}},}\ }\href {\doibase 10.1103/PhysRevB.94.014427} {\bibfield  {journal} {\bibinfo  {journal} {Phys. Rev. B}\ }\textbf {\bibinfo {volume} {94}},\ \bibinfo {pages} {014427} (\bibinfo {year} {2016})}\BibitemShut {NoStop}%
\bibitem [{\citenamefont {Lv}\ \emph {et~al.}(2018)\citenamefont {Lv}, \citenamefont {Kally}, \citenamefont {Zhang}, \citenamefont {Lee}, \citenamefont {Jamali}, \citenamefont {Samarth},\ and\ \citenamefont {Wang}}]{Lv_NatCommun2018}%
  \BibitemOpen
  \bibfield  {author} {\bibinfo {author} {\bibfnamefont {Y.}~\bibnamefont {Lv}}, \bibinfo {author} {\bibfnamefont {J.}~\bibnamefont {Kally}}, \bibinfo {author} {\bibfnamefont {D.}~\bibnamefont {Zhang}}, \bibinfo {author} {\bibfnamefont {J.S.}\ \bibnamefont {Lee}}, \bibinfo {author} {\bibfnamefont {M.}~\bibnamefont {Jamali}}, \bibinfo {author} {\bibfnamefont {N.}~\bibnamefont {Samarth}}, \ and\ \bibinfo {author} {\bibfnamefont {J.-P.}\ \bibnamefont {Wang}},\ }\bibfield  {title} {\enquote {\bibinfo {title} {Unidirectional spin-{H}all and {R}ashba-{E}delstein magnetoresistance in topological insulator-ferromagnet layer heterostructures},}\ }\href {\doibase 10.1038/s41467-017-02491-3} {\bibfield  {journal} {\bibinfo  {journal} {Nat. Commun.}\ }\textbf {\bibinfo {volume} {9}},\ \bibinfo {pages} {111} (\bibinfo {year} {2018})}\BibitemShut {NoStop}%
\bibitem [{\citenamefont {Lv}\ \emph {et~al.}(2022)\citenamefont {Lv}, \citenamefont {Kally}, \citenamefont {Liu}, \citenamefont {Quarterman}, \citenamefont {Pillsbury}, \citenamefont {Kirby}, \citenamefont {Grutter}, \citenamefont {Sahu}, \citenamefont {Borchers}, \citenamefont {Wu}, \citenamefont {Samarth},\ and\ \citenamefont {Wang}}]{Lv_ApplPhysRev2022}%
  \BibitemOpen
  \bibfield  {author} {\bibinfo {author} {\bibfnamefont {Y.}~\bibnamefont {Lv}}, \bibinfo {author} {\bibfnamefont {J.}~\bibnamefont {Kally}}, \bibinfo {author} {\bibfnamefont {T.}~\bibnamefont {Liu}}, \bibinfo {author} {\bibfnamefont {P.}~\bibnamefont {Quarterman}}, \bibinfo {author} {\bibfnamefont {T.}~\bibnamefont {Pillsbury}}, \bibinfo {author} {\bibfnamefont {B.~J.}\ \bibnamefont {Kirby}}, \bibinfo {author} {\bibfnamefont {A.~J.}\ \bibnamefont {Grutter}}, \bibinfo {author} {\bibfnamefont {P.}~\bibnamefont {Sahu}}, \bibinfo {author} {\bibfnamefont {J.~A.}\ \bibnamefont {Borchers}}, \bibinfo {author} {\bibfnamefont {M.}~\bibnamefont {Wu}}, \bibinfo {author} {\bibfnamefont {N.}~\bibnamefont {Samarth}}, \ and\ \bibinfo {author} {\bibfnamefont {J.-P.}\ \bibnamefont {Wang}},\ }\bibfield  {title} {\enquote {\bibinfo {title} {Large unidirectional spin {H}all and {R}ashba--{E}delstein magnetoresistance in topological insulator/magnetic insulator heterostructures},}\ }\href {\doibase 10.1063/5.0073976} {\bibfield
  {journal} {\bibinfo  {journal} {Appl. Phys. Rev.}\ }\textbf {\bibinfo {volume} {9}},\ \bibinfo {pages} {011406} (\bibinfo {year} {2022})}\BibitemShut {NoStop}%
\bibitem [{\citenamefont {Singh}\ \emph {et~al.}(2024)\citenamefont {Singh}, \citenamefont {Raman},\ and\ \citenamefont {Mohanta}}]{JSingh_PRB2024}%
  \BibitemOpen
  \bibfield  {author} {\bibinfo {author} {\bibfnamefont {J.}~\bibnamefont {Singh}}, \bibinfo {author} {\bibfnamefont {K.~V.}\ \bibnamefont {Raman}}, \ and\ \bibinfo {author} {\bibfnamefont {N.}~\bibnamefont {Mohanta}},\ }\bibfield  {title} {\enquote {\bibinfo {title} {Anisotropic planar {H}all effects in {B}i$_{2}${S}e$_{3}$/{E}u{S} interfaces: Deciphering the role of proximity-induced spin canting and topological spin texture},}\ }\href {\doibase 10.1103/PhysRevB.110.125133} {\bibfield  {journal} {\bibinfo  {journal} {Phys. Rev. B}\ }\textbf {\bibinfo {volume} {110}},\ \bibinfo {pages} {125133} (\bibinfo {year} {2024})}\BibitemShut {NoStop}%
\bibitem [{\citenamefont {Suri}\ \emph {et~al.}(2024)\citenamefont {Suri}, \citenamefont {Sasmal}, \citenamefont {Bhardwaj}, \citenamefont {Singh}, \citenamefont {Mundlia}, \citenamefont {Mishra}, \citenamefont {Mohanta},\ and\ \citenamefont {Raman}}]{Dhavala_PRB2024}%
  \BibitemOpen
  \bibfield  {author} {\bibinfo {author} {\bibfnamefont {D.}~\bibnamefont {Suri}}, \bibinfo {author} {\bibfnamefont {S.}~\bibnamefont {Sasmal}}, \bibinfo {author} {\bibfnamefont {A.}~\bibnamefont {Bhardwaj}}, \bibinfo {author} {\bibfnamefont {J.}~\bibnamefont {Singh}}, \bibinfo {author} {\bibfnamefont {S.}~\bibnamefont {Mundlia}}, \bibinfo {author} {\bibfnamefont {A.}~\bibnamefont {Mishra}}, \bibinfo {author} {\bibfnamefont {N.}~\bibnamefont {Mohanta}}, \ and\ \bibinfo {author} {\bibfnamefont {K.~V.}\ \bibnamefont {Raman}},\ }\bibfield  {title} {\enquote {\bibinfo {title} {Emergence of planar topological {H}all anisotropy in {B}i$_{2}$({S}e,{T}e)$_{3}$ by proximity-induced spin-canted state of the heisenberg ferromagnetic insulator {E}u{S}},}\ }\href {\doibase 10.1103/PhysRevB.110.134433} {\bibfield  {journal} {\bibinfo  {journal} {Phys. Rev. B}\ }\textbf {\bibinfo {volume} {110}},\ \bibinfo {pages} {134433} (\bibinfo {year} {2024})}\BibitemShut {NoStop}%
\bibitem [{\citenamefont {Mohanta}\ \emph {et~al.}(2017)\citenamefont {Mohanta}, \citenamefont {Kampf},\ and\ \citenamefont {Kopp}}]{Mohanta_SciRep2017}%
  \BibitemOpen
  \bibfield  {author} {\bibinfo {author} {\bibfnamefont {N.}~\bibnamefont {Mohanta}}, \bibinfo {author} {\bibfnamefont {A.}~\bibnamefont {Kampf}}, \ and\ \bibinfo {author} {\bibfnamefont {T.}~\bibnamefont {Kopp}},\ }\bibfield  {title} {\enquote {\bibinfo {title} {Emergent momentum-space skyrmion texture on the surface of topological insulators},}\ }\href {\doibase https://doi.org/10.1038/srep45664} {\bibfield  {journal} {\bibinfo  {journal} {Sci. Rep.}\ }\textbf {\bibinfo {volume} {7}},\ \bibinfo {pages} {45664} (\bibinfo {year} {2017})}\BibitemShut {NoStop}%
\bibitem [{\citenamefont {Wang}\ \emph {et~al.}(2015)\citenamefont {Wang}, \citenamefont {Deorani}, \citenamefont {Banerjee}, \citenamefont {Koirala}, \citenamefont {Brahlek}, \citenamefont {Oh},\ and\ \citenamefont {Yang}}]{Wang_PRL2015}%
  \BibitemOpen
  \bibfield  {author} {\bibinfo {author} {\bibfnamefont {Y.}~\bibnamefont {Wang}}, \bibinfo {author} {\bibfnamefont {P.}~\bibnamefont {Deorani}}, \bibinfo {author} {\bibfnamefont {K.}~\bibnamefont {Banerjee}}, \bibinfo {author} {\bibfnamefont {N.}~\bibnamefont {Koirala}}, \bibinfo {author} {\bibfnamefont {M.}~\bibnamefont {Brahlek}}, \bibinfo {author} {\bibfnamefont {S.}~\bibnamefont {Oh}}, \ and\ \bibinfo {author} {\bibfnamefont {H.}~\bibnamefont {Yang}},\ }\bibfield  {title} {\enquote {\bibinfo {title} {Topological surface states originated spin-orbit torques in {B}i$_{2}${S}e$_{3}$},}\ }\href {\doibase 10.1103/PhysRevLett.114.257202} {\bibfield  {journal} {\bibinfo  {journal} {Phys. Rev. Lett.}\ }\textbf {\bibinfo {volume} {114}},\ \bibinfo {pages} {257202} (\bibinfo {year} {2015})}\BibitemShut {NoStop}%
\bibitem [{\citenamefont {Garate}\ and\ \citenamefont {Franz}(2010)}]{Garate_PRL2010}%
  \BibitemOpen
  \bibfield  {author} {\bibinfo {author} {\bibfnamefont {I.}~\bibnamefont {Garate}}\ and\ \bibinfo {author} {\bibfnamefont {M.}~\bibnamefont {Franz}},\ }\bibfield  {title} {\enquote {\bibinfo {title} {Inverse spin-galvanic effect in the interface between a topological insulator and a ferromagnet},}\ }\href {\doibase 10.1103/PhysRevLett.104.146802} {\bibfield  {journal} {\bibinfo  {journal} {Phys. Rev. Lett.}\ }\textbf {\bibinfo {volume} {104}},\ \bibinfo {pages} {146802} (\bibinfo {year} {2010})}\BibitemShut {NoStop}%
\bibitem [{\citenamefont {Chen}(2020)}]{Chen_PRB2020}%
  \BibitemOpen
  \bibfield  {author} {\bibinfo {author} {\bibfnamefont {W.}~\bibnamefont {Chen}},\ }\bibfield  {title} {\enquote {\bibinfo {title} {Spin torque and persistent currents caused by percolation of topological surface states},}\ }\href {\doibase 10.1103/PhysRevB.102.144442} {\bibfield  {journal} {\bibinfo  {journal} {Phys. Rev. B}\ }\textbf {\bibinfo {volume} {102}},\ \bibinfo {pages} {144442} (\bibinfo {year} {2020})}\BibitemShut {NoStop}%
\bibitem [{\citenamefont {Ghosh}\ and\ \citenamefont {Manchon}(2018)}]{Ghosh_PRB2018}%
  \BibitemOpen
  \bibfield  {author} {\bibinfo {author} {\bibfnamefont {S.}~\bibnamefont {Ghosh}}\ and\ \bibinfo {author} {\bibfnamefont {A.}~\bibnamefont {Manchon}},\ }\bibfield  {title} {\enquote {\bibinfo {title} {Spin-orbit torque in a three-dimensional topological insulator--ferromagnet heterostructure: Crossover between bulk and surface transport},}\ }\href {\doibase 10.1103/PhysRevB.97.134402} {\bibfield  {journal} {\bibinfo  {journal} {Phys. Rev. B}\ }\textbf {\bibinfo {volume} {97}},\ \bibinfo {pages} {134402} (\bibinfo {year} {2018})}\BibitemShut {NoStop}%
\bibitem [{\citenamefont {Ezawa}(2024)}]{Ezawa_PRB2025}%
  \BibitemOpen
  \bibfield  {author} {\bibinfo {author} {\bibfnamefont {M.}~\bibnamefont {Ezawa}},\ }\bibfield  {title} {\enquote {\bibinfo {title} {Detecting the n\'eel vector of altermagnets in heterostructures with a topological insulator and a crystalline valley-edge insulator},}\ }\href {\doibase 10.1103/PhysRevB.109.245306} {\bibfield  {journal} {\bibinfo  {journal} {Phys. Rev. B}\ }\textbf {\bibinfo {volume} {109}},\ \bibinfo {pages} {245306} (\bibinfo {year} {2024})}\BibitemShut {NoStop}%
\bibitem [{\citenamefont {Ma}\ and\ \citenamefont {Jia}(2024{\natexlab{a}})}]{Hai-Yang_PRB2024}%
  \BibitemOpen
  \bibfield  {author} {\bibinfo {author} {\bibfnamefont {H.-Y.}\ \bibnamefont {Ma}}\ and\ \bibinfo {author} {\bibfnamefont {J.-F.}\ \bibnamefont {Jia}},\ }\bibfield  {title} {\enquote {\bibinfo {title} {Altermagnetic topological insulator and the selection rules},}\ }\href {\doibase 10.1103/PhysRevB.110.064426} {\bibfield  {journal} {\bibinfo  {journal} {Phys. Rev. B}\ }\textbf {\bibinfo {volume} {110}},\ \bibinfo {pages} {064426} (\bibinfo {year} {2024}{\natexlab{a}})}\BibitemShut {NoStop}%
\bibitem [{\citenamefont {Ma}\ and\ \citenamefont {Jia}(2024{\natexlab{b}})}]{Ma_Quantumfront2024}%
  \BibitemOpen
  \bibfield  {author} {\bibinfo {author} {\bibfnamefont {H.-Y.}\ \bibnamefont {Ma}}\ and\ \bibinfo {author} {\bibfnamefont {J.-F.}\ \bibnamefont {Jia}},\ }\bibfield  {title} {\enquote {\bibinfo {title} {Altermagnetic topological insulator with {C}-paired spin-valley locking},}\ }\href {\doibase 10.1007/s44214-024-00070-4} {\bibfield  {journal} {\bibinfo  {journal} {Quantum Front.}\ }\textbf {\bibinfo {volume} {3}},\ \bibinfo {pages} {22} (\bibinfo {year} {2024}{\natexlab{b}})}\BibitemShut {NoStop}%
\bibitem [{\citenamefont {Pan}\ \emph {et~al.}(2015)\citenamefont {Pan}, \citenamefont {Wu}, \citenamefont {Liu},\ and\ \citenamefont {Yang}}]{Hui_SciRep2015}%
  \BibitemOpen
  \bibfield  {author} {\bibinfo {author} {\bibfnamefont {H.}~\bibnamefont {Pan}}, \bibinfo {author} {\bibfnamefont {M.}~\bibnamefont {Wu}}, \bibinfo {author} {\bibfnamefont {Y.}~\bibnamefont {Liu}}, \ and\ \bibinfo {author} {\bibfnamefont {S.~A.}\ \bibnamefont {Yang}},\ }\bibfield  {title} {\enquote {\bibinfo {title} {Electric control of topological phase transitions in {D}irac semimetal thin films},}\ }\href {\doibase 10.1038/srep14639} {\bibfield  {journal} {\bibinfo  {journal} {Sci. Rep.}\ }\textbf {\bibinfo {volume} {5}} (\bibinfo {year} {2015}),\ 10.1038/srep14639}\BibitemShut {NoStop}%
\bibitem [{\citenamefont {Sødequist}\ and\ \citenamefont {Olsen}(2024)}]{Joachim_APL2024}%
  \BibitemOpen
  \bibfield  {author} {\bibinfo {author} {\bibfnamefont {J.}~\bibnamefont {Sødequist}}\ and\ \bibinfo {author} {\bibfnamefont {T.}~\bibnamefont {Olsen}},\ }\bibfield  {title} {\enquote {\bibinfo {title} {Two-dimensional altermagnets from high throughput computational screening: Symmetry requirements, chiral magnons, and spin-orbit effects},}\ }\href {\doibase 10.1063/5.0198285} {\bibfield  {journal} {\bibinfo  {journal} {Appl. Phys. Lett.}\ }\textbf {\bibinfo {volume} {124}},\ \bibinfo {pages} {182409} (\bibinfo {year} {2024})}\BibitemShut {NoStop}%
\bibitem [{\citenamefont {Milivojević}\ \emph {et~al.}(2024)\citenamefont {Milivojević}, \citenamefont {Orozović}, \citenamefont {Picozzi}, \citenamefont {Gmitra},\ and\ \citenamefont {Stavrić}}]{Milivojević_2DMater2024}%
  \BibitemOpen
  \bibfield  {author} {\bibinfo {author} {\bibfnamefont {M.}~\bibnamefont {Milivojević}}, \bibinfo {author} {\bibfnamefont {M.}~\bibnamefont {Orozović}}, \bibinfo {author} {\bibfnamefont {S.}~\bibnamefont {Picozzi}}, \bibinfo {author} {\bibfnamefont {M.}~\bibnamefont {Gmitra}}, \ and\ \bibinfo {author} {\bibfnamefont {S.}~\bibnamefont {Stavrić}},\ }\bibfield  {title} {\enquote {\bibinfo {title} {Interplay of altermagnetism and weak ferromagnetism in two-dimensional {R}u{F}$_4$},}\ }\href {\doibase 10.1088/2053-1583/ad4c73} {\bibfield  {journal} {\bibinfo  {journal} {2D Mater.}\ }\textbf {\bibinfo {volume} {11}},\ \bibinfo {pages} {035025} (\bibinfo {year} {2024})}\BibitemShut {NoStop}%
\bibitem [{\citenamefont {David}\ \emph {et~al.}(2019)\citenamefont {David}, \citenamefont {Rakyta}, \citenamefont {Korm\'anyos},\ and\ \citenamefont {Burkard}}]{David_PRB2019}%
  \BibitemOpen
  \bibfield  {author} {\bibinfo {author} {\bibfnamefont {A.}~\bibnamefont {David}}, \bibinfo {author} {\bibfnamefont {P.}~\bibnamefont {Rakyta}}, \bibinfo {author} {\bibfnamefont {A.}~\bibnamefont {Korm\'anyos}}, \ and\ \bibinfo {author} {\bibfnamefont {G.}~\bibnamefont {Burkard}},\ }\bibfield  {title} {\enquote {\bibinfo {title} {Induced spin-orbit coupling in twisted graphene--transition metal dichalcogenide heterobilayers: {T}wistronics meets spintronics},}\ }\href {\doibase 10.1103/PhysRevB.100.085412} {\bibfield  {journal} {\bibinfo  {journal} {Phys. Rev. B}\ }\textbf {\bibinfo {volume} {100}},\ \bibinfo {pages} {085412} (\bibinfo {year} {2019})}\BibitemShut {NoStop}%
\bibitem [{\citenamefont {Xiao}\ \emph {et~al.}(2010)\citenamefont {Xiao}, \citenamefont {Chang},\ and\ \citenamefont {Niu}}]{Xiao_RMP2010}%
  \BibitemOpen
  \bibfield  {author} {\bibinfo {author} {\bibfnamefont {D.}~\bibnamefont {Xiao}}, \bibinfo {author} {\bibfnamefont {M.-C.}\ \bibnamefont {Chang}}, \ and\ \bibinfo {author} {\bibfnamefont {Q.}~\bibnamefont {Niu}},\ }\bibfield  {title} {\enquote {\bibinfo {title} {{B}erry phase effects on electronic properties},}\ }\href {\doibase 10.1103/RevModPhys.82.1959} {\bibfield  {journal} {\bibinfo  {journal} {Rev. Mod. Phys.}\ }\textbf {\bibinfo {volume} {82}},\ \bibinfo {pages} {1959--2007} (\bibinfo {year} {2010})}\BibitemShut {NoStop}%
\bibitem [{\citenamefont {Mahan}(2000)}]{Mahan_ManyParticle}%
  \BibitemOpen
  \bibfield  {author} {\bibinfo {author} {\bibfnamefont {G.~D.}\ \bibnamefont {Mahan}},\ }\href@noop {} {\emph {\bibinfo {title} {Many-Particle Physics}}},\ \bibinfo {edition} {3rd}\ ed.\ (\bibinfo  {publisher} {Springer US},\ \bibinfo {address} {Boston, MA},\ \bibinfo {year} {2000})\BibitemShut {NoStop}%
\bibitem [{\citenamefont {Fedchenko}\ \emph {et~al.}(2024)\citenamefont {Fedchenko}, \citenamefont {Min\'a\v{r}}, \citenamefont {Akashdeep}, \citenamefont {D'Souza}, \citenamefont {Vasilyev}, \citenamefont {Tkach}, \citenamefont {Odenbreit}, \citenamefont {Nguyen}, \citenamefont {Kutnyakhov}, \citenamefont {Wind}, \citenamefont {Wenthaus}, \citenamefont {Scholz}, \citenamefont {Rossnagel}, \citenamefont {Hoesch}, \citenamefont {Aeschlimann}, \citenamefont {Stadtm\"uller}, \citenamefont {Kl\"aui}, \citenamefont {Sch\"onhense}, \citenamefont {Jakob}, \citenamefont {{Šmejkal}}, \citenamefont {Sinova},\ and\ \citenamefont {Elmers}}]{Fedchenko_SciAdv2024}%
  \BibitemOpen
  \bibfield  {author} {\bibinfo {author} {\bibfnamefont {O.}~\bibnamefont {Fedchenko}}, \bibinfo {author} {\bibfnamefont {J.}~\bibnamefont {Min\'a\v{r}}}, \bibinfo {author} {\bibfnamefont {A.}~\bibnamefont {Akashdeep}}, \bibinfo {author} {\bibfnamefont {S.~W.}\ \bibnamefont {D'Souza}}, \bibinfo {author} {\bibfnamefont {D.}~\bibnamefont {Vasilyev}}, \bibinfo {author} {\bibfnamefont {O.}~\bibnamefont {Tkach}}, \bibinfo {author} {\bibfnamefont {L.}~\bibnamefont {Odenbreit}}, \bibinfo {author} {\bibfnamefont {Q.~L.}\ \bibnamefont {Nguyen}}, \bibinfo {author} {\bibfnamefont {D.}~\bibnamefont {Kutnyakhov}}, \bibinfo {author} {\bibfnamefont {N.}~\bibnamefont {Wind}}, \bibinfo {author} {\bibfnamefont {L.}~\bibnamefont {Wenthaus}}, \bibinfo {author} {\bibfnamefont {M.}~\bibnamefont {Scholz}}, \bibinfo {author} {\bibfnamefont {K.}~\bibnamefont {Rossnagel}}, \bibinfo {author} {\bibfnamefont {M.}~\bibnamefont {Hoesch}}, \bibinfo {author} {\bibfnamefont {M.}~\bibnamefont {Aeschlimann}}, \bibinfo {author} {\bibfnamefont
  {B.}~\bibnamefont {Stadtm\"uller}}, \bibinfo {author} {\bibfnamefont {M.}~\bibnamefont {Kl\"aui}}, \bibinfo {author} {\bibfnamefont {G.}~\bibnamefont {Sch\"onhense}}, \bibinfo {author} {\bibfnamefont {G.}~\bibnamefont {Jakob}}, \bibinfo {author} {\bibfnamefont {L.}~\bibnamefont {{Šmejkal}}}, \bibinfo {author} {\bibfnamefont {J.}~\bibnamefont {Sinova}}, \ and\ \bibinfo {author} {\bibfnamefont {H.-J.}\ \bibnamefont {Elmers}},\ }\bibfield  {title} {\enquote {\bibinfo {title} {Observation of time-reversal symmetry breaking in the band structure of altermagnetic {R}u{O}$_2$},}\ }\href {\doibase 10.1126/sciadv.adj4883} {\bibfield  {journal} {\bibinfo  {journal} {Sci. Adv.}\ }\textbf {\bibinfo {volume} {10}},\ \bibinfo {pages} {eadj4883} (\bibinfo {year} {2024})}\BibitemShut {NoStop}%
\bibitem [{\citenamefont {Guo}\ \emph {et~al.}(2024)\citenamefont {Guo}, \citenamefont {Zhang}, \citenamefont {Zengtai}, \citenamefont {Jiang}, \citenamefont {Jiang}, \citenamefont {Wu}, \citenamefont {Dong}, \citenamefont {Xu}, \citenamefont {He}, \citenamefont {He}, \citenamefont {Huang}, \citenamefont {Du}, \citenamefont {Zhang}, \citenamefont {Wu}, \citenamefont {Han}, \citenamefont {Shao}, \citenamefont {Yu},\ and\ \citenamefont {Wu}}]{Guo_AdvSci2024}%
  \BibitemOpen
  \bibfield  {author} {\bibinfo {author} {\bibfnamefont {Y.}~\bibnamefont {Guo}}, \bibinfo {author} {\bibfnamefont {J.}~\bibnamefont {Zhang}}, \bibinfo {author} {\bibfnamefont {Z.}~\bibnamefont {Zengtai}}, \bibinfo {author} {\bibfnamefont {Y.-Y.}\ \bibnamefont {Jiang}}, \bibinfo {author} {\bibfnamefont {L.}~\bibnamefont {Jiang}}, \bibinfo {author} {\bibfnamefont {C.}~\bibnamefont {Wu}}, \bibinfo {author} {\bibfnamefont {J.}~\bibnamefont {Dong}}, \bibinfo {author} {\bibfnamefont {X.}~\bibnamefont {Xu}}, \bibinfo {author} {\bibfnamefont {W.}~\bibnamefont {He}}, \bibinfo {author} {\bibfnamefont {B.}~\bibnamefont {He}}, \bibinfo {author} {\bibfnamefont {Z.}~\bibnamefont {Huang}}, \bibinfo {author} {\bibfnamefont {L.}~\bibnamefont {Du}}, \bibinfo {author} {\bibfnamefont {G.}~\bibnamefont {Zhang}}, \bibinfo {author} {\bibfnamefont {K.}~\bibnamefont {Wu}}, \bibinfo {author} {\bibfnamefont {X.}~\bibnamefont {Han}}, \bibinfo {author} {\bibfnamefont {D.-F.}\ \bibnamefont {Shao}}, \bibinfo {author} {\bibfnamefont
  {G.}~\bibnamefont {Yu}}, \ and\ \bibinfo {author} {\bibfnamefont {H.}~\bibnamefont {Wu}},\ }\bibfield  {title} {\enquote {\bibinfo {title} {Direct and inverse spin splitting effects in altermagnetic {R}u{O}$_2$},}\ }\href {\doibase 10.1002/advs.202400967} {\bibfield  {journal} {\bibinfo  {journal} {Adv. Sci}\ }\textbf {\bibinfo {volume} {11}},\ \bibinfo {pages} {2400967} (\bibinfo {year} {2024})}\BibitemShut {NoStop}%
\bibitem [{\citenamefont {Osumi}\ \emph {et~al.}(2024)\citenamefont {Osumi}, \citenamefont {Souma}, \citenamefont {Aoyama}, \citenamefont {Yamauchi}, \citenamefont {Honma}, \citenamefont {Nakayama}, \citenamefont {Takahashi}, \citenamefont {Ohgushi},\ and\ \citenamefont {Sato}}]{Osumi_PRB2024}%
  \BibitemOpen
  \bibfield  {author} {\bibinfo {author} {\bibfnamefont {T.}~\bibnamefont {Osumi}}, \bibinfo {author} {\bibfnamefont {S.}~\bibnamefont {Souma}}, \bibinfo {author} {\bibfnamefont {T.}~\bibnamefont {Aoyama}}, \bibinfo {author} {\bibfnamefont {K.}~\bibnamefont {Yamauchi}}, \bibinfo {author} {\bibfnamefont {A.}~\bibnamefont {Honma}}, \bibinfo {author} {\bibfnamefont {K.}~\bibnamefont {Nakayama}}, \bibinfo {author} {\bibfnamefont {T.}~\bibnamefont {Takahashi}}, \bibinfo {author} {\bibfnamefont {K.}~\bibnamefont {Ohgushi}}, \ and\ \bibinfo {author} {\bibfnamefont {T.}~\bibnamefont {Sato}},\ }\bibfield  {title} {\enquote {\bibinfo {title} {Observation of a giant band splitting in altermagnetic {M}n{T}e},}\ }\href {\doibase 10.1103/PhysRevB.109.115102} {\bibfield  {journal} {\bibinfo  {journal} {Phys. Rev. B}\ }\textbf {\bibinfo {volume} {109}},\ \bibinfo {pages} {115102} (\bibinfo {year} {2024})}\BibitemShut {NoStop}%
\bibitem [{\citenamefont {Amin}\ \emph {et~al.}(2024)\citenamefont {Amin}, \citenamefont {Dal~Din}, \citenamefont {Golias}, \citenamefont {Niu}, \citenamefont {Zakharov}, \citenamefont {Fromage}, \citenamefont {Fields}, \citenamefont {Heywood}, \citenamefont {Cousins}, \citenamefont {Maccherozzi}, \citenamefont {Krempaský}, \citenamefont {Dil}, \citenamefont {Kriegner}, \citenamefont {Kiraly}, \citenamefont {Campion}, \citenamefont {Rushforth}, \citenamefont {Edmonds}, \citenamefont {Dhesi}, \citenamefont {Šmejkal}, \citenamefont {Jungwirth},\ and\ \citenamefont {Wadley}}]{Amin_Nature2024}%
  \BibitemOpen
  \bibfield  {author} {\bibinfo {author} {\bibfnamefont {O.~J.}\ \bibnamefont {Amin}}, \bibinfo {author} {\bibfnamefont {A.}~\bibnamefont {Dal~Din}}, \bibinfo {author} {\bibfnamefont {E.}~\bibnamefont {Golias}}, \bibinfo {author} {\bibfnamefont {Y.}~\bibnamefont {Niu}}, \bibinfo {author} {\bibfnamefont {A.}~\bibnamefont {Zakharov}}, \bibinfo {author} {\bibfnamefont {S.~C.}\ \bibnamefont {Fromage}}, \bibinfo {author} {\bibfnamefont {C.~J.~B.}\ \bibnamefont {Fields}}, \bibinfo {author} {\bibfnamefont {S.~L.}\ \bibnamefont {Heywood}}, \bibinfo {author} {\bibfnamefont {R.~B.}\ \bibnamefont {Cousins}}, \bibinfo {author} {\bibfnamefont {F.}~\bibnamefont {Maccherozzi}}, \bibinfo {author} {\bibfnamefont {J.}~\bibnamefont {Krempaský}}, \bibinfo {author} {\bibfnamefont {J.~H.}\ \bibnamefont {Dil}}, \bibinfo {author} {\bibfnamefont {D.}~\bibnamefont {Kriegner}}, \bibinfo {author} {\bibfnamefont {B.}~\bibnamefont {Kiraly}}, \bibinfo {author} {\bibfnamefont {R.~P.}\ \bibnamefont {Campion}}, \bibinfo {author}
  {\bibfnamefont {A.~W.}\ \bibnamefont {Rushforth}}, \bibinfo {author} {\bibfnamefont {K.~W.}\ \bibnamefont {Edmonds}}, \bibinfo {author} {\bibfnamefont {S.~S.}\ \bibnamefont {Dhesi}}, \bibinfo {author} {\bibfnamefont {L.}~\bibnamefont {Šmejkal}}, \bibinfo {author} {\bibfnamefont {T.}~\bibnamefont {Jungwirth}}, \ and\ \bibinfo {author} {\bibfnamefont {P.}~\bibnamefont {Wadley}},\ }\bibfield  {title} {\enquote {\bibinfo {title} {Nanoscale imaging and control of altermagnetism in {M}n{T}e},}\ }\href {\doibase 10.1038/s41586-024-08234-x} {\bibfield  {journal} {\bibinfo  {journal} {Nature}\ }\textbf {\bibinfo {volume} {636}},\ \bibinfo {pages} {348--353} (\bibinfo {year} {2024})}\BibitemShut {NoStop}%
\bibitem [{\citenamefont {Reimers}\ \emph {et~al.}(2024)\citenamefont {Reimers}, \citenamefont {N{\v o}ky}, \citenamefont {G{\"o}bel}, \citenamefont {Johansson}, \citenamefont {Jaiswal}, \citenamefont {Warnicke}, \citenamefont {Zhang}, \citenamefont {Felser},\ and\ \citenamefont {Reiss}}]{Reimers_NatCommun2024}%
  \BibitemOpen
  \bibfield  {author} {\bibinfo {author} {\bibfnamefont {S.}~\bibnamefont {Reimers}}, \bibinfo {author} {\bibfnamefont {J.}~\bibnamefont {N{\v o}ky}}, \bibinfo {author} {\bibfnamefont {B.}~\bibnamefont {G{\"o}bel}}, \bibinfo {author} {\bibfnamefont {A.}~\bibnamefont {Johansson}}, \bibinfo {author} {\bibfnamefont {S.}~\bibnamefont {Jaiswal}}, \bibinfo {author} {\bibfnamefont {P.}~\bibnamefont {Warnicke}}, \bibinfo {author} {\bibfnamefont {H.}~\bibnamefont {Zhang}}, \bibinfo {author} {\bibfnamefont {C.}~\bibnamefont {Felser}}, \ and\ \bibinfo {author} {\bibfnamefont {G.}~\bibnamefont {Reiss}},\ }\bibfield  {title} {\enquote {\bibinfo {title} {Direct observation of altermagnetic band splitting in {C}r{S}b thin films},}\ }\href {\doibase 10.1038/s41467-024-46476-5} {\bibfield  {journal} {\bibinfo  {journal} {Nat. Commun.}\ }\textbf {\bibinfo {volume} {15}},\ \bibinfo {pages} {4647} (\bibinfo {year} {2024})}\BibitemShut {NoStop}%
\bibitem [{\citenamefont {Nan}\ \emph {et~al.}(2019)\citenamefont {Nan}, \citenamefont {Anderson}, \citenamefont {Gibbons}, \citenamefont {Hwang}, \citenamefont {Campbell}, \citenamefont {Zhou}, \citenamefont {Dong}, \citenamefont {Kim}, \citenamefont {Shao}, \citenamefont {Paudel}, \citenamefont {Reynolds}, \citenamefont {Wang}, \citenamefont {Sun}, \citenamefont {Tsymbal}, \citenamefont {Choi}, \citenamefont {Rzchowski}, \citenamefont {Kim}, \citenamefont {Ralph},\ and\ \citenamefont {Eom}}]{Nan_PNAS2019}%
  \BibitemOpen
  \bibfield  {author} {\bibinfo {author} {\bibfnamefont {T.}~\bibnamefont {Nan}}, \bibinfo {author} {\bibfnamefont {T.~J.}\ \bibnamefont {Anderson}}, \bibinfo {author} {\bibfnamefont {J.}~\bibnamefont {Gibbons}}, \bibinfo {author} {\bibfnamefont {K.}~\bibnamefont {Hwang}}, \bibinfo {author} {\bibfnamefont {N.}~\bibnamefont {Campbell}}, \bibinfo {author} {\bibfnamefont {H.}~\bibnamefont {Zhou}}, \bibinfo {author} {\bibfnamefont {Y.~Q.}\ \bibnamefont {Dong}}, \bibinfo {author} {\bibfnamefont {G.~Y.}\ \bibnamefont {Kim}}, \bibinfo {author} {\bibfnamefont {D.~F.}\ \bibnamefont {Shao}}, \bibinfo {author} {\bibfnamefont {T.~R.}\ \bibnamefont {Paudel}}, \bibinfo {author} {\bibfnamefont {N.}~\bibnamefont {Reynolds}}, \bibinfo {author} {\bibfnamefont {X.~J.}\ \bibnamefont {Wang}}, \bibinfo {author} {\bibfnamefont {N.~X.}\ \bibnamefont {Sun}}, \bibinfo {author} {\bibfnamefont {E.~Y.}\ \bibnamefont {Tsymbal}}, \bibinfo {author} {\bibfnamefont {S.~Y.}\ \bibnamefont {Choi}}, \bibinfo {author} {\bibfnamefont {M.~S.}\
  \bibnamefont {Rzchowski}}, \bibinfo {author} {\bibfnamefont {Y.~B.}\ \bibnamefont {Kim}}, \bibinfo {author} {\bibfnamefont {D.~C.}\ \bibnamefont {Ralph}}, \ and\ \bibinfo {author} {\bibfnamefont {C.~B.}\ \bibnamefont {Eom}},\ }\bibfield  {title} {\enquote {\bibinfo {title} {Anisotropic spin-orbit torque generation in epitaxial {S}r{I}r{O}$_3$ by symmetry design},}\ }\href {\doibase 10.1073/pnas.1812822116} {\bibfield  {journal} {\bibinfo  {journal} {Proc. Natl. Acad. Sci.}\ }\textbf {\bibinfo {volume} {116}},\ \bibinfo {pages} {16186--16191} (\bibinfo {year} {2019})}\BibitemShut {NoStop}%
\bibitem [{\citenamefont {Ali}\ \emph {et~al.}(2014)\citenamefont {Ali}, \citenamefont {Xiong}, \citenamefont {Flynn}, \citenamefont {Tao}, \citenamefont {Gibson}, \citenamefont {Schoop}, \citenamefont {Liang}, \citenamefont {Haldolaarachchige}, \citenamefont {Hirschberger}, \citenamefont {Ong},\ and\ \citenamefont {Cava}}]{Ali_WTe2_Nature2014}%
  \BibitemOpen
  \bibfield  {author} {\bibinfo {author} {\bibfnamefont {M.~N.}\ \bibnamefont {Ali}}, \bibinfo {author} {\bibfnamefont {J.}~\bibnamefont {Xiong}}, \bibinfo {author} {\bibfnamefont {S.}~\bibnamefont {Flynn}}, \bibinfo {author} {\bibfnamefont {J.}~\bibnamefont {Tao}}, \bibinfo {author} {\bibfnamefont {Q.~D.}\ \bibnamefont {Gibson}}, \bibinfo {author} {\bibfnamefont {L.~M.}\ \bibnamefont {Schoop}}, \bibinfo {author} {\bibfnamefont {T.}~\bibnamefont {Liang}}, \bibinfo {author} {\bibfnamefont {N.}~\bibnamefont {Haldolaarachchige}}, \bibinfo {author} {\bibfnamefont {M.}~\bibnamefont {Hirschberger}}, \bibinfo {author} {\bibfnamefont {N.~P.}\ \bibnamefont {Ong}}, \ and\ \bibinfo {author} {\bibfnamefont {R.~J.}\ \bibnamefont {Cava}},\ }\bibfield  {title} {\enquote {\bibinfo {title} {Large, non-saturating magnetoresistance in {WT}e$_2$},}\ }\href {\doibase 10.1038/nature13763} {\bibfield  {journal} {\bibinfo  {journal} {Nature}\ }\textbf {\bibinfo {volume} {514}},\ \bibinfo {pages} {205--208} (\bibinfo {year}
  {2014})}\BibitemShut {NoStop}%
\bibitem [{\citenamefont {Zhang}\ \emph {et~al.}(2009)\citenamefont {Zhang}, \citenamefont {Liu}, \citenamefont {Qi}, \citenamefont {Dai}, \citenamefont {Fang},\ and\ \citenamefont {Zhang}}]{Zhang_NatPhys2009}%
  \BibitemOpen
  \bibfield  {author} {\bibinfo {author} {\bibfnamefont {H.}~\bibnamefont {Zhang}}, \bibinfo {author} {\bibfnamefont {C.-X.}\ \bibnamefont {Liu}}, \bibinfo {author} {\bibfnamefont {X-L}\ \bibnamefont {Qi}}, \bibinfo {author} {\bibfnamefont {X.}~\bibnamefont {Dai}}, \bibinfo {author} {\bibfnamefont {Z.}~\bibnamefont {Fang}}, \ and\ \bibinfo {author} {\bibfnamefont {S.-C.}\ \bibnamefont {Zhang}},\ }\bibfield  {title} {\enquote {\bibinfo {title} {Topological insulators in {B}i$_2${S}e$_3$, {B}i$_2${T}e$_3$ and {S}b$_2${T}e$_3$ with a single {D}irac cone on the surface.}}\ }\href {\doibase https://doi.org/10.1038/nphys1270} {\bibfield  {journal} {\bibinfo  {journal} {Nat. Phys.}\ ,\ \bibinfo {pages} {438--442}} (\bibinfo {year} {2009})}\BibitemShut {NoStop}%
\bibitem [{\citenamefont {Liu}\ \emph {et~al.}(2010)\citenamefont {Liu}, \citenamefont {Qi}, \citenamefont {Zhang}, \citenamefont {Dai}, \citenamefont {Fang},\ and\ \citenamefont {Zhang}}]{Liu_modelTI_PRB2010}%
  \BibitemOpen
  \bibfield  {author} {\bibinfo {author} {\bibfnamefont {C.-X.}\ \bibnamefont {Liu}}, \bibinfo {author} {\bibfnamefont {X.-L.}\ \bibnamefont {Qi}}, \bibinfo {author} {\bibfnamefont {H.}~\bibnamefont {Zhang}}, \bibinfo {author} {\bibfnamefont {X.}~\bibnamefont {Dai}}, \bibinfo {author} {\bibfnamefont {Z.}~\bibnamefont {Fang}}, \ and\ \bibinfo {author} {\bibfnamefont {S.-C.}\ \bibnamefont {Zhang}},\ }\bibfield  {title} {\enquote {\bibinfo {title} {Model hamiltonian for topological insulators},}\ }\href {\doibase 10.1103/PhysRevB.82.045122} {\bibfield  {journal} {\bibinfo  {journal} {Phys. Rev. B}\ }\textbf {\bibinfo {volume} {82}},\ \bibinfo {pages} {045122} (\bibinfo {year} {2010})}\BibitemShut {NoStop}%
\bibitem [{\citenamefont {Gupta}\ \emph {et~al.}(2014)\citenamefont {Gupta}, \citenamefont {Jalil},\ and\ \citenamefont {Liang}}]{Gupta_SciRep_2014}%
  \BibitemOpen
  \bibfield  {author} {\bibinfo {author} {\bibfnamefont {G.}~\bibnamefont {Gupta}}, \bibinfo {author} {\bibfnamefont {M.}~\bibnamefont {Jalil}}, \ and\ \bibinfo {author} {\bibfnamefont {G.}~\bibnamefont {Liang}},\ }\bibfield  {title} {\enquote {\bibinfo {title} {Evaluation of mobility in thin {B}i$_2${S}e$_3$ topological insulator for prospects of local electrical interconnects},}\ }\href {\doibase 10.1038/srep06838} {\bibfield  {journal} {\bibinfo  {journal} {Sci. Rep.}\ }\textbf {\bibinfo {volume} {4}},\ \bibinfo {pages} {6838} (\bibinfo {year} {2014})}\BibitemShut {NoStop}%
\end{thebibliography}%

\end{document}